\documentclass[review,12pt]{elsarticle}
\journal{Journal of Electrostatics}
\usepackage[overload]{empheq} 
\usepackage{amsmath}
\usepackage{amssymb}
\usepackage[left=2.5cm, right=2.5cm, top=2.5cm]{geometry}
\usepackage{booktabs}
\usepackage{dcolumn}
\usepackage{bm}
\usepackage{verbatim}
\usepackage[utf8]{inputenc}
\usepackage[T1]{fontenc}
\usepackage{appendix}
\usepackage{url}
\usepackage{ulem}

\usepackage{graphicx}
\usepackage{threeparttable} 
\usepackage{float}
\usepackage{subfig} 
\usepackage{placeins}
\usepackage[colorlinks=true,linkcolor=black,anchorcolor=black,citecolor=black,filecolor=black,menucolor=black,runcolor=black,urlcolor=black]{hyperref} 
\usepackage{cleveref}

\usepackage{listings}
\usepackage{array}

\usepackage{multirow}
\usepackage{setspace}

\newcommand{\diverg}{\nabla\cdot}

\newcommand{\parder}[2]{\frac{\partial #1}{\partial #2}}

\usepackage{tikz}
\newcommand{\full}{\tikz \draw[fill=black] (0,0) circle (0.12cm);}
\newcommand{\emptyc}{\tikz \draw[fill=white] (0,0) circle (0.12cm);}

\newcommand{\patIII}[3]{#1\ #2\ #3}
\newcommand{\patV}[5]{#1\ #2\ #3\ #4\ #5}

\usepackage{algorithm}
\usepackage{algorithmic}
\makeatletter
\def\@email#1#2{%
 \endgroup
 \patchcmd{\titleblock@produce}
  {\frontmatter@RRAPformat}
  {\frontmatter@RRAPformat{\produce@RRAP{*#1\href{mailto:#2}{#2}}}\frontmatter@RRAPformat}
  {}{}
}%
\makeatother
\begin{document}
\begin{frontmatter}

\title{A Macroscopic Model of Bipolar Corona Discharges for Ionic Thrusters Simulation}
\author[1]{G.\ Cali\`o}
\author[1]{F.G.\ Fischetti}
\author[1]{L.\ Valdettaro}
\author[1]{C.\ de Falco}
\author[1]{P.\ Barbante\corref{cor1}}
\affiliation[1]{organization={Politecnico di Milano, MOX, Department of Mathematics}, city={Milano}, country={Italy}}
\cortext[cor1]{Corresponding author: paolo.barbante@polimi.it}
\date{}
\begin{abstract}
Recent studies on Electro-Aero-Dynamic (EAD) thrusters have demonstrated that optimized, multi-stage, electrode arrangements can lead to substantial performance improvements. Within this context, numerical models can play a crucial role in the design process. Existing models can be broadly classified into full plasma models, that account for the chemical kinetics in the ionization region and reduced models, that rely on simplifying assumptions, \textit{e.g.}\ Kaptzov's hypothesis, to avoid the computation of the ionization region. Full plasma models, despite their accuracy, are computationally demanding, especially when coupled with the Navier-Stokes equations. Reduced models are computationally cheaper and therefore are well suited for the design process. Compact design EAD thrusters that use multiple stages with alternating polarity have recently appeared in literature. Such a configuration allows smaller inter-stage distances, because the emitter(s) of a stage cannot be ignited by the collector(s) of the previous one, as no voltage difference is applied between them. In this work, we introduce a reduced bipolar corona discharge model coupled to the Navier-Stokes equations. A numerical simulation code is implemented based on the high performance finite element deal.II library. Our model can successfully simulate multi-stage thrusters with alternating polarity configurations and automatically detect the ignition pattern, when a configuration with multiple emitters is considered. Validation is performed against previous experimental and numerical campaigns, demonstrating the good predictive capabilities of our model.     
\end{abstract}

\end{frontmatter}

\section{Introduction}
Electro-Aero-Dynamic (EAD) thrusters have been extensively investigated both numerically and experimentally in recent years~\cite{Belan21,Belan24,Nature,Barrett20, Terenzi25,Usuelli25}. Usually they operate in the corona discharge regime, which occurs when a sufficiently high voltage is applied to an electrode, the \textit{emitter}, placed opposite to a grounded \textit{collector}. Once the electric field around the emitter exceeds the onset threshold, $E_{on}$, ionization by electron collisions is triggered. Ions are accelerated toward the collector, moving in the \textit{drift region} under the action of the Coulomb force. The momentum exchanged by ions with the background gas generates the \textit{ionic wind}, producing a thrust in the opposite direction. Depending on the polarity, the discharge can be classified as either positive or negative: for positive corona the drift region is primarily populated by positive ions, whereas for negative corona the charge carriers are predominantly negative ions.\\
Positive corona is employed in the majority of recent studies~\cite{Belan21,Belan24,Nature,Barrett20, Terenzi25,Usuelli25}, but negative corona is expected to become increasingly important. Reference~\cite{GomezVega2024} studied multi-stage configurations that outperformed the thrust of single-stage geometries. Multi-stage thrusters can be arranged either by repeating stages with the same polarity (usually positive), or by alternating polarity across successive stages~\cite{Barrett15,GomezVega2023}. In the first case, particular care must be taken in setting the inter-stage distance, which should be large enough to prevent the onset of ionic counter wind, \textit{i.e.}\ the drift of ions from emitter(s) of a stage to the collector(s) of the previous one, that reduces the engine's thrust. The arrangement with alternating polarity (\textit{e.g.}\ even stages have negative polarity, while odd stages positive one~\cite{GomezVega2024}) allows for reduced inter-stage distances, because emitter(s) of a stage and collector(s) of the previous one share the same polarity, therefore ionic counter wind is not possible. This latter configuration highlights the need for bipolar models in numerical simulations.\\
Another aspect relevant for EAD thrusters is related to the non-trivial inception (\textit{i.e.}\ the beginning of ionization) patterns that arise in multiple emitters configurations. Reference~\cite{MultiInception} investigated this problem both numerically and experimentally, considering wire-to-wire setups with various numbers of emitters and a single collector. The multi-Kaptzov model~\cite{Coseru} used by Ref.~\cite{MultiInception} was unable to automatically identify which emitters were active
and, as a work-around, an ad-hoc trial and error strategy was used. 
Despite the legitimacy of this approach, automatic detection of ignited electrodes remains a highly desirable feature for numerical models.\\
A possible modeling strategy is to resort to full plasma models~\cite{Parent,Villa,Morrow}, coupled with kinetic schemes of variable complexity to reproduce the chemical processes underlying the discharge phenomenon. Despite their accuracy, these models are computationally demanding, especially when coupled with the Navier-Stokes equations. Monrolin et al.~\cite{MonPlour} developed a two-domain model, that solves the kinetic problem with Townsend reactions in the ionization region only, thus reducing the computational cost. Reduced models represent another alternative that can further decrease the computational burden. We can refer to them as \textit{macroscopic} models because they solve for the drift region only, neglecting the plasma kinetics in the thin ionization region, that needs to be replaced by a suitable boundary condition~\cite{Coseru,Cagnoni,Calio}. In Refs.~\cite{Cagnoni,Calio} an injection, or emission, law is prescribed, linking the number density of generated ions with the magnitude of the electric field at the emitter. Since the emission of ions is related to the local electric field, the inception pattern can be detected automatically~\cite{Calio}. However, the approach of Ref.~\cite{Calio} has been developed for positive polarity discharges only.
In this work, we aim to extend this approach to alternating polarity discharges.\\
The paper is organized as follows. In Section~\ref{sec: model} we describe the physical models and the underlying assumptions. In Section~\ref{sec: NumericalMethods} we give some details on the numerical methods. In Section~\ref{sec: Validation} we perform a validation against experimental studies~\cite{GomezVega2024} for a single stage thruster. In Section~\ref{sec: Multistage} we analyze the effect of recombination between ions in multi-stage thrusters and we compare our results against a two stage experimental configuration. In Section~\ref{sec:perfeval} we analyze performance characterization in terms of the inter-stage distance. Finally, in Section~\ref{sec: MultipleEmitterInception} we demonstrate that the ignition pattern is automatically detected when multiple emitters are considered. 

\section{Physical Models}\label{sec: model}
\begin{figure}
    \centering
    \includegraphics[width=0.8\linewidth]{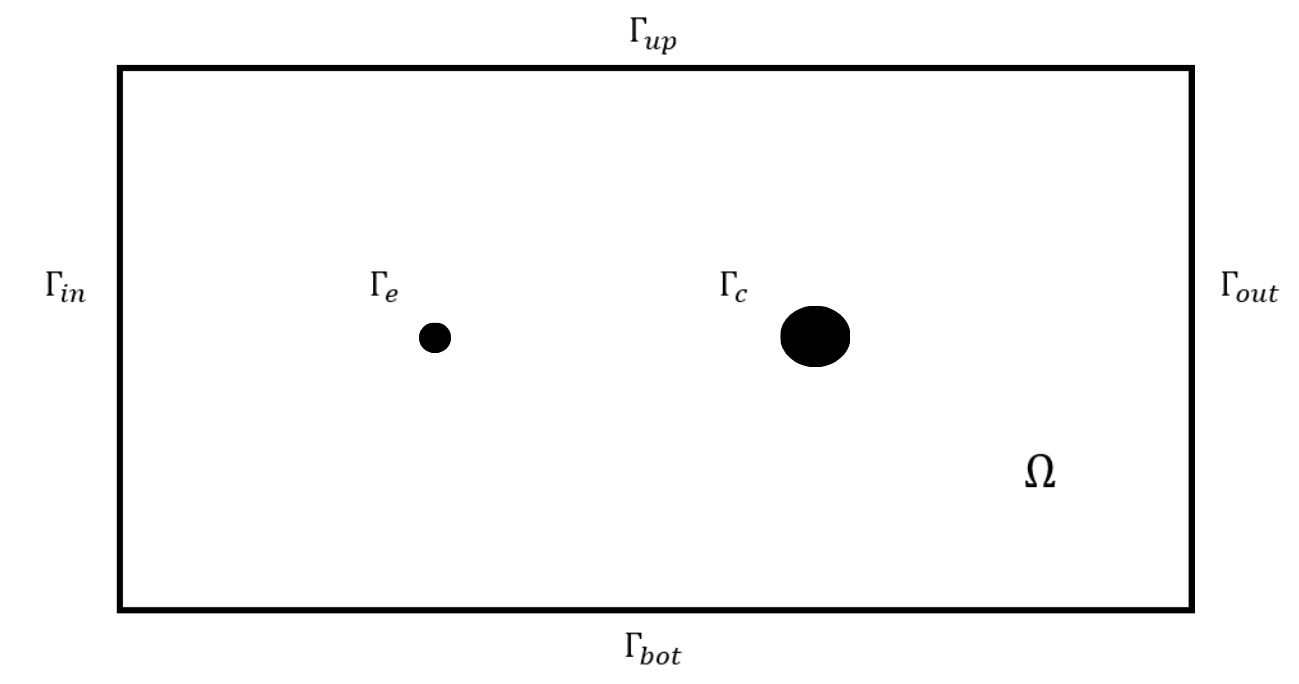}
    \caption{Reference computational domain.}
    \label{fig: domain}
\end{figure}

In our approach we couple a drift-diffusion model for the corona discharge with a Navier-Stokes model for the flow-field induced by the ionic wind. The combined model is applied to a computational domain $\Omega$ sketched in Fig.~\ref{fig: domain}.
$\Gamma_{in}$ and $\Gamma_{out}$ are the fluid inlet and outlet parts of the domain respectively. $\Gamma_{up}$ and $\Gamma_{bot}$ are the upper and lower boundaries of the domain.
$\Gamma_e = \partial\Omega_e$ is the surface of the emitter(s), $\Gamma_c = \partial\Omega_c$ is the surface of the collector(s). 

\subsection{Bipolar Corona Model}
\label{sec:coronamodel}

We consider two macro-species, positive ions, denoted by subscript $p$ and negative ions, denoted by subscript $n$.
The ion number densities are computed by fluid equations under the drift diffusion approximation, whereas the electric field is modeled by the Poisson equation: 
\begin{equation}
\left\{
\begin{aligned}
     &-\nabla \cdot \left(\varepsilon \nabla \varphi\right) = q (n_p-n_n)\\
     & \nabla \cdot\left(- D_p \nabla n_p - \mu_p\ n_p\ \nabla \varphi\right)= -R_i\left(n_n\ n_p -n_{min}^2\right) \\
     & \nabla \cdot\left(- D_n \nabla n_n + \mu_n\ n_n\ \nabla \varphi\right)= -R_i\left(n_n\ n_p -n_{min}^2\right) 
     \label{eq: DD}
\end{aligned}
\right.
\end{equation}
$\varphi$ is the electric potential and the electric field is defined as $\vec{E}=-\nabla \varphi$ under the electrostatic hypothesis. $n_p$ and $n_n$ are the number density of positive and negative ions respectively. The parameter $n_{min}$ prescribes a minimum seed of ions~\cite{Calio}. $\varepsilon$ is the dielectric constant of air; $q$ the fundamental charge; $\mu_p$ and $D_p$ the mobility and diffusion coefficients of positive ions, $\mu_n$ and $D_n$ the mobility and diffusion coefficients of negative ions. Mobility and diffusion coefficients are related by the Einstein-Smoluchowski equation 
    $D = \mu k_{B} T/{q}  = \mu V_{th}$, 
where $k_B$ is the Boltzmann constant, $T$ is temperature of the ions, assumed to be in thermal equilibrium with the neutral gas and $V_{th}$ is the thermal potential.
The current densities of positive and negative ions are defined respectively as:
$\vec{J}_p = - D_p \nabla n_p - \mu_p n_p \nabla \varphi$ and $\vec{J}_n = D_n \nabla n_n - \mu_n n_n \nabla \varphi$. The source term $R_i\left(n_n n_p -n_{min}^2\right)$ takes into account possible recombination (neutralization) reactions between positive and negative ions~\cite{Bipolar}, where $R_i$ is a constant. The influence of this term on discharge computations is discussed in Sec.~\ref{sec: Recombination}.\\
We remark that a production term for ions is absent in Eqs.~\ref{eq: DD}. The ionization region is very close to the active electrode (emitter), where the electric field is high enough to trigger the ionization processes. In our reduced model the ionization region is not solved for, but it is instead modeled by a Dirichlet boundary condition that acts as an injection law~\cite{Cagnoni,Calio} for ions at the emitter's surface. The boundary condition at the emitter takes the form:
\begin{equation}
\left\{
\begin{aligned}
    n_p(\vec{E}\cdot \vec{n})&=f_p(\vec{E}\cdot \vec{n})=  \max \left[n_{min},\;n_{ref,p}\exp\left(\frac{\vec{E}\cdot \vec{n}-\Bar{E}_{on,p}}{E_{ref,p}}\right)\right]\\
    n_n(\vec{E}\cdot \vec{n})&=f_n(\vec{E}\cdot \vec{n})= \max\left[n_{min},\;n_{ref,n}\exp\left(\frac{-\vec{E}\cdot \vec{n}-\Bar{E}_{on,n}}{E_{ref,n}}\right)  \right]
\end{aligned}
    \right.\label{eq: BC_ion}
\end{equation}
where $\vec{n}$ is the outgoing normal vector. The boundary condition prescribes the amount of positive or negative ions injected into the drift region as a function of the local value of the electric field at the emitter's surface. We assume the same mathematical form of the ignition law for negative and positive corona discharges. The computation of parameters $n_{min}$, $n_{ref,p}$, $\bar{E}_{on,p}$ and $E_{ref,p}$ in Eq.~\ref{eq: BC_ion} has been discussed in Ref.~\cite{Calio} for the case of positive ions. The parameters $n_{ref,n}$, $\Bar{E}_{on,n}$, $E_{ref,n}$ for negative ions can differ from the corresponding ones for positive ions. In this contribution we assume $n_{ref,n}=n_{ref,p}=n_{ref}$, $E_{ref,n}=E_{ref,p}=E_{ref}$; on the contrary we use $\Bar{E}_{on,p}=\Bar{E}_{on}$, and $\Bar{E}_{on,n}= 0.9\,\Bar{E}_{on,p}$. This choice is consistent with the typical behavior of negative discharges that tend to ignite before the positive ones, as it was observed experimentally in Ref.~\cite{GomezVega2024}. 
At the collector(s) we impose $n_p=n_n=n_{min}$ and a zero current condition at the exterior boundaries of the domain ($\Gamma_{in}$, $\Gamma_{out}$, $\Gamma_{bot}$, $\Gamma_{up}$). For the electric potential a fixed value is imposed at the electrodes (the collector(s) being grounded) and null normal derivative at the exterior boundaries. Emitter's voltage is positive for positive corona and negative for negative one.
Therefore the complete set of boundary conditions for Equations~\ref{eq: DD} reads:
\begin{equation}
    \left\{
    \begin{aligned}
        &\varphi = \pm V_a ;\quad\quad n_p=f_p(\vec{E} \cdot \vec{n}); \quad n_n=f_n(\vec{E}\cdot \vec{n}) &  \Gamma_e\\ 
    &\varphi = 0 ;\quad\quad  n_p =n_n=n_{min} & \Gamma_c\\ 
    &\nabla \varphi \cdot \vec{n}=0;\quad \vec{J}_p\cdot \vec{n}=0;\quad\quad\quad\quad \vec{J}_n\cdot \vec{n}=0 & \partial\Omega \setminus  (\Gamma_e \cup \Gamma_c)
    \end{aligned}\label{eq: BC_DD}
    \right.
\end{equation}
We refer to Equations~\ref{eq: DD}, with their boundary conditions (Eqs.~\ref{eq: BC_DD}), as the “corona problem”. The choice of a stationary model for corona discharge is motivated by the fact that the characteristic time scale of the discharge phenomenon is much smaller than the time scale of the fluid flow. Thus, it is reasonable to assume that the corona discharge reaches a steady state almost instantaneously compared to the fluid dynamics.

\subsection{Fluid Model}
\label{sec:fluidmodel}

The fluid problem, namely the motion of the plasma mixture, is modeled by the incompressible Navier-Stokes equations:
\begin{equation}
    \left\{
    \begin{aligned}
        \rho_f\parder{\vec{u}}{t}+\rho_f(\vec{u}\cdot\nabla)\vec{u} + \nabla p=\mu_f\Delta\vec{u}+\vec{f}_{E} \\
        \diverg{\vec{u}}=0
    \end{aligned}\label{eq: NS}
    \right.
\end{equation}
Rigorously the plasma total density $\rho$ is given by
$\rho = {m_p n_p} + {m_n n_n} + \rho_f$, where $\rho_f$ is the density of the neutral species.
However, in a weakly ionized plasma like a corona discharge, the contribution of ionized species to the total density is negligible compared to the density of neutral species and $\rho$ can be reduced to $\rho \approx \rho_f$.
In our model we consider the corona problem one-way coupled with the Navier-Stokes equations: only the effect of the corona on the airflow is taken into account. The electrostatic volume force $\vec{f}_E$ due to the corona discharge that acts on the fluid is computed as:
\begin{equation}
    \vec{f}_{E}=q (n_p-n_n) \vec{E}
    \label{eq:ehdforce}
\end{equation}
The boundary conditions for the fluid problem read: 
\begin{subequations}
    \label{eq: NS_BC}
    \begin{align}[left=\empheqlbrace]
        &\vec{u}=0 & \left(\Gamma_{e}\cup \Gamma_{c}\right) \label{eq: NS_BC_noslip}\\
        &\mathbb{T}\cdot\vec{n}+\frac{1}{2}\left(\vec{u}\cdot\vec{n}\right)_{\delta}\ \vec{u}=0\ & \left(\Gamma_{in} \cup \Gamma_{out}\right) \label{eq: NS_BC_RDDN}\\
        &\vec{u}\cdot\vec{n}=0 &   (\Gamma_{bot}\cup\Gamma_{up}) \label{eq: NS_BC_freeslip}
    \end{align}
\end{subequations}
where $\mathbb{T}_{ij}=\mu_f\frac{\partial u_i}{\partial x_j}-p\delta_{ij}$ is an "artificial" stress tensor.
The boundary condition applied at the inlet $\Gamma_{in}$ and at the outlet $\Gamma_{out}$ is the \textit{regularized directional-do-nothing} (RDDN)~\cite{Nogueira2025} condition, where
\begin{equation}
\left(\vec{u}\cdot\vec{n}\right)_{\delta} = \frac{1}{2}\left[\left(\vec{u}\cdot\vec{n}\right)\tanh\left(\frac{\vec{u}\cdot\vec{n}}{\delta}\right)-\left(\vec{u}\cdot\vec{n}\right)+\delta\right]\label{eq: RDDN}
\end{equation}
and $0 < \delta \ll 1$ is a regularization parameter; in this contribution $\delta=10^{-5}$.
This boundary condition allows us to prescribe an open boundary for both inflow and outflow, without knowing in advance the amount of incoming flow. This approach eliminates the need to prescribe a small inlet velocity, as done in Ref.~\cite{Picella}. From a theoretical point of view, it can be shown that the RDDN condition has the desirable property of improving the flow stability compared to the simple do-nothing condition~\cite{Nogueira2025}.\\

\subsection{Current and thrust computation}

The current per unit length $I$ (whose units are $\mathrm{A/m}$) is computed at the electrodes:
\begin{equation}
    I =  q \int_{\Gamma_e} \left( \vec{J}_p + \vec{J}_n \right)\cdot \vec{n}\, d s = - q \int_{\Gamma_c} \left( \vec{J}_p + \vec{J}_n \right)\cdot \vec{n}\, d s
    \label{eq:current}
\end{equation}
The electrostatic thrust $\vec{T}_{E}$ is the force applied on the electrodes by the electrostatic field only:
\begin{equation}
    \vec{T}_{E}=\int_{\Omega}\vec{f}_{E}\,dV = \int_{\Omega}q (n_p-n_n) \vec{E}\,dV
    \label{eq:Tehd}
\end{equation}
The effective, or electro-aerodynamic, thrust $\vec{T}_{EAD}$ is the total force applied to the electrodes; it takes into account also the effects of the flow-field and can be computed as: 
\begin{equation}
    \vec{T}_{EAD}=\vec{T}_{E}-\vec{D}=\int_{\Omega} \vec{f}_{E}\,dV+\int_{\left(\Gamma_{c}\,\cup\, \Gamma_{e}\right)}\mathbb{T}^f\cdot \vec{n}\,ds
\end{equation}
$\mathbb{T}_{ij}^f = \mu_f\left(\frac{\partial u_i}{\partial x_j}+\frac{\partial u_j}{\partial x_i}\right)-p\delta_{ij}$ is the true fluid stress tensor.
Because the computations are carried out on two-dimensional configurations, $\vec{T}_E$ and $\vec{T}_{EAD}$ have the dimension of a force per unit length ($\mathrm{N/m}$).
The effective thrust $\vec{T}_{EAD}$ can also be computed with an integral balance of momentum over the domain $\Omega$:
\begin{equation}
    \vec{T}_{EAD}=\int_{\Gamma_{in}} p \vec{n}ds+ \int_{\Gamma_{out}} p\vec{n} ds +\int_{\Gamma_{in}} \rho_f \vec{u}(\vec{u}\cdot \vec{n}) ds+\int_{\Gamma_{out}} \rho_f \vec{u}(\vec{u}\cdot \vec{n}) ds \label{eq:T_momentum}
    \end{equation}
The contribution of the upper and bottom boundaries is neglected because of the boundary condition Eq.~\ref{eq: NS_BC_freeslip}. The viscous term on the inlet and outlet boundaries is neglected too.
For the applications of interest in this paper, we compute only the horizontal component of $\vec{T}_{E}$ and $\vec{T}_{EAD}$ in the discussion of the results, being the vertical component negligible. 

\section{Numerical Methods}
\label{sec: NumericalMethods}

The numerical solution of the models presented in Secs.~\ref{sec:coronamodel} and~\ref{sec:fluidmodel} is carried out for two-dimensional grids composed of quadrilateral elements. The numerical methods are implemented on top of the deal.II library~\cite{dealii}, an open source finite element library that provides a wide range of tools for the implementation of finite element methods, including mesh handling, linear algebra operations, and linear solvers. Meshes are generated with the GMSH library~\cite{gmsh}. Through the deal.II interface, local refinements are handled by the p4est library~\cite{BursteddeWilcoxGhattas11}, the data structures by PETSc~\cite{petsc-web-page,petsc-efficient} and the linear solver by MUMPS \cite{MUMPS:1,MUMPS:2}.

\subsection{Dimensionless Formulation of the Corona Problem}

The governing equations are made dimensionless with the reference parameters shown in Table~\ref{tab: reference_quantities}:
\begin{table}[H]
\centering
\begin{tabular}{l c r}
  \toprule
  \textbf{Symbol} & \textbf{Value}       & \textbf{S.I. Units} \\
    $\overline{L}$  &  $r_c$                 & $[\mathrm{m}]$  \\
    $\overline{V}$  &  $V_a$ & $[\mathrm{V}]$  \\
    $\overline{n}_p$  &  $\overline{L}^{-2} q^{-1}\ \varepsilon \overline{V}$ 
                    &  $[\mathrm{m}^{-3}]$                     \\
    $\overline{n}_n$  &  $\overline{L}^{-2} q^{-1}\ \varepsilon \overline{V}$ 
                    &  $[\mathrm{m}^{-3}]$                     \\
    $\overline{E}$  &  $\overline{L}^{-1}  \overline{V}$ 
                    &  $[\mathrm{m}^{-1}\ \mathrm{V}]$   \\
    $\overline{R}_i^p$  &  $\overline{n}_p^{-1}\overline{L}^{-2} D_p$  
                    &  $[\mathrm{m}^3\ \mathrm{s}^{-1}]$   \\
    $\overline{R}_i^n$  & $ \overline{n}_n^{-1}\overline{L}^{-2} D_n$  
                    &  $[\mathrm{m}^3\ \mathrm{s}^{-1}]$   \\
\end{tabular}
\caption{Reference Quantities}\label{tab: reference_quantities}
\end{table}
This non-dimensionalization strategy was already used in Ref.~\cite{Coseru}, with the difference that in our contribution an additional recombination term is present in the r.h.s.\ of the drift-diffusion equations.
The P\'eclet number $Pe=\overline{V}/V_{th}$ and the non-dimensional recombination coefficients $R^p=R_i/\overline{R}_i^p$ and $R^n=R_i/\overline{R}_i^n$ are the only non-dimensional quantities appearing in the equations. We remark that the non-dimensional quantities $R^p$ and $R^n$ are in principle different, because diffusion coefficients are not equal for positive and negative ions, due to differences in the mobilities.
Denoting with the symbol $\widehat{\left(\, \cdot\, \right)}$ the non-dimensional variables, the dimensionless formulation reads:
\begin{equation}
\left\{
\begin{aligned}
     &-\widehat{\nabla} \cdot \left(  \widehat{\nabla} \widehat{\varphi} \right) = \left( \widehat{n}_p - \widehat{n}_n \right)\\
     & -\widehat{\nabla} \cdot \left( \frac{1}{Pe} \widehat{\nabla} \widehat{n}_p + \widehat{n}_p\ \widehat{\nabla} \widehat{\varphi} \right)=-\frac{R^p}{Pe}\ (\widehat{n}_p\ \widehat{n}_n- \widehat{n}_{min}^2) \\
     & -\widehat{\nabla} \cdot \left( \frac{1}{Pe} \widehat{\nabla} \widehat{n}_n - \widehat{n}_n\ \widehat{\nabla} \widehat{\varphi} \right)=-\frac{R^n}{Pe}\ (\widehat{n}_p\ \widehat{n}_n -\widehat n_{min}^2) 
     \label{eq: DD_adim}
\end{aligned}
\right.
\end{equation}

\subsection{Discretization of the Corona Problem}

\begin{figure}[H]
    \centering
    \includegraphics[width=0.5\linewidth]{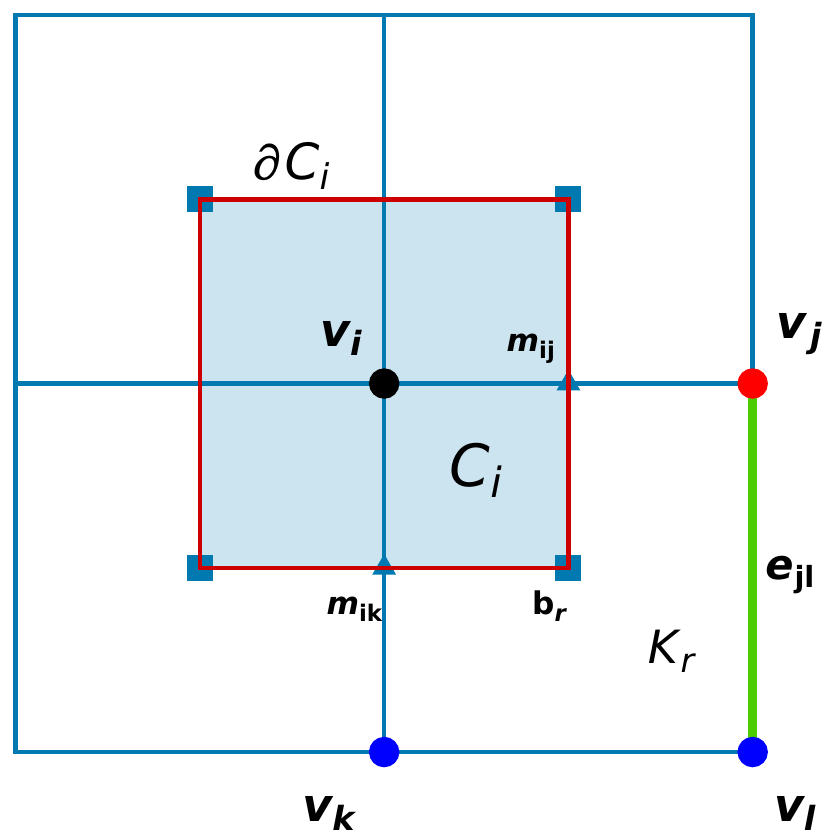}
    \caption{Dual control volume $C_i$ associated with the vertex $v_i$ of the primal mesh.}
    \label{fig:control_volume}
\end{figure}

A scheme for the spatial discretization of Eqs.~\ref{eq: DD_adim} needs to satisfy the following properties: stability in the strong drift regime, \textit{i.e.}\ when drift velocity induced by the electric field dominates over diffusion flux and cell-by-cell conservation of positive and negative current densities. Moreover, for the applications considered in this work, the discretization should be applicable to general quadrilateral meshes.\\
Control Volume Finite Element Method (CVFEM) is an approach in which the primal Galerkin method is combined with the local conservation properties of the Finite Volume Method, with the advantage that CVFEM does not need a topologically dual grid~\cite{only_CVFEM}. CVFEM approximates the unknowns using nodal basis functions, as the primal Galerkin method, onto a weak form of the equations.\\
The computational domain $\Omega$ is divided into a set of primal cells $\mathrm{K}_r$, each with barycenter $\mathbf{b}_r$. The vertices of the primal mesh are denoted by $\mathbf{v}_i$; $\mathbf{e}_{ij}$ is the edge that connects vertices $\mathbf{v}_i$ and $\mathbf{v}_j$. The midpoint $\mathbf{m}_{ij}$ and the length $h_{ij}$ of $\mathbf{e}_{ij}$ are defined as:
$\mathbf{m}_{ij}=(\mathbf{v}_i+\mathbf{v}_j)/2$ and
$h_{ij}=|\mathbf{v}_i-\mathbf{v}_j|$.
The dual control volume $C_i$, associated with the vertex $\mathbf{v}_i$, is constructed by connecting the barycenters of the primal cells adjacent to $\mathbf{v}_i$ with the midpoint of the edges incident to vertex $\mathbf{v}_i$.
Figure~\ref{fig:control_volume} shows a sketch of the primal grid and of the dual control volume for
a two-dimensional grid.
One can remark that the dual control volume $C_i$ naturally splits into four sub-control volumes in two dimensions.
$V(\Omega)$ is the set of all primal mesh vertices, $E(\Omega)$ is the set of all edges, while $E(\mathbf{v}_i)$ is the set of edges incident to vertex $\mathbf{v}_i$. For a cell ${K}_r$ having $\mathbf{v}_i$ as one of its vertices, $C_i^r$ denotes the portion of the dual control volume $C_i$ contained in $K_r$. Each control volume $C_i$ has area $\vert C_i\vert $, while the sub-control volume $C_i^r$ on $K_r$ has area $\vert C_i^r\vert $; the contour of $C_i$ is $\partial C_i$. $\mathrm{K}(\mathbf{v}_i)$ denotes the set of cells having $\mathbf{v}_i$ as a vertex.
The weak form of Eqs.~\ref{eq: DD_adim} is obtained by applying the Divergence Theorem to every dual control volume $C_i$. If we write Eqs.~\ref{eq: DD_adim} with the contracted notation $\diverg \vec{J} = R$, the integral weak form reads:
\begin{equation}
    \int_{\partial C_i}\vec{J} \cdot \vec{n}\,dS= \int_{C_i} R \,dV \quad \forall \mathbf{v}_i\in V(\Omega)
\end{equation}
When CVFEM is applied to drift-dominated problems, some sort of stabilization is needed~\cite{stabilized_only_CVFEM} for the numerical approximation of the flux $\int_{\partial C_i}\vec{J} \cdot \vec{n}\,dS$. In a standard CVFEM approach the current $\vec{J}$ is approximated using the values computed at the vertices of the primal grid (the nodal current)~\cite{CVFEM}.
The main idea of the stabilized CVFEM-SG method is to replace the nodal current with one-dimensional Scharfetter-Gummel currents~\cite{schar69} computed along the edges $\mathbf{e}_{ij}$ of each primal mesh element~\cite{CVFEM}.
Stability is enhanced with an edge element lifting of the one-dimensional Scharfetter-Gummel edge currents into curl-conforming currents. 
The elemental curl-conforming current is defined as:
\begin{equation}
\vec{J}_E=\sum_{e_{ij}\in E(\Omega)} h_{ij} J_{ij} \vec{W}_{ij}\label{eq: Je}
\end{equation}
In this expression, $\vec{W}_{ij}$ are the N\'ed\'elec~\cite{Nedelec} basis functions and $J_{ij}$ is the one-dimensional Scharfetter-Gummel expression of the current along the edge $\mathbf{e}_{ij}$:
\begin{equation} 
J_{ij}= \frac{D}{h_{ij}}[n_j B(-2 a_{ij})-n_i B(2 a_{ij})]
\label{eq: SG_1D_formula}
\end{equation}
where $a_{ij}=-q(\varphi_j-\varphi_i)/(2 k_B T)$
and $B(x)=x/(\exp(x)-1)$ is the Bernoulli function.
Substituting Eq.~\ref{eq: SG_1D_formula} into Eq.~\ref{eq: Je} we have a computable expression of the elemental current~\cite{CVFEM}. The discretized equation finally is:
\begin{equation}
    \int_{\partial C_i}\vec{J_E} \cdot \vec{n}\,dS= \int_{C_i} R \,dV \quad \forall \mathbf{v}_i\in V(\Omega)
\end{equation}
Further details on the CVFEM-SG numerical approximation are given in Appendix~\ref{sec: discretized_newton}.

\subsection{Solution of the Non-linear Corona Problem}

We solve the non linear system arising from the numerical approximation of Eqs.~\ref{eq: DD_adim} with a global Newton method. An equation is added for the normal component of the electric field at the emitter $E_{\Gamma_e}=\vec{E}_{\Gamma_e}\cdot\hat{n}$. The additional equation for the normal electric field at the emitter boundary, as required by the injection law of Eqs.~\ref{eq: BC_ion}, facilitates the assembly of the Jacobian matrix.
The computation of the discretized residuals and of the Jacobian matrix is described Appendix~\ref{sec: discretized_newton}.\\
To improve convergence of the Newton iteration, we apply a numerical continuation method using
the emitter voltage as the continuation parameter; we note in passing that, because of the scaling procedure adopted here, this is equivalent to perform a continuation with respect to the inverse P\'eclet number $1/Pe$. Each inner iteration has been solved with the sparse direct solver MUMPS \cite{MUMPS:1, MUMPS:2}. 
In this work stopping criteria for the Newton iteration have been chosen as follows:
\begin{equation}
    \begin{aligned}
        &\left\lVert \frac{\vert\delta \varphi^{(k)}\vert}{\vert\varphi^{(k)}\vert + 1} \right\lVert_{\infty} < tol\ ;  \
        &\left\lVert \frac{\vert \delta n_p^{(k)}\vert}{\vert n_p^{(k)} + n_{ref}\vert}\right\lVert_{\infty} < tol\ ;  \
        &\left\lVert \frac{\vert\delta n_n^{(k)}\vert}{\vert n_n^{(k)} + n_{ref}\vert}\right\lVert_{\infty} < tol\ ;  \
        &\left\lVert \frac{\vert\delta E_{\Gamma_{e}}^{(k)}\vert}{\vert E_{\Gamma_{e}}^{(k)}\vert + E_{ref}}\right\lVert_{\infty} < tol 
    \end{aligned}
    \notag
\end{equation}
The initial guess is computed as follows. First, a constant charge distribution is assigned in the entire domain $n_{p}^{(0)} = n_{n}^{(0)} = n_{min}$. The potential $\varphi_0^{(0)}$ is computed by solving the Poisson equation. Finally the initial electric field at the emitter(s) is computed $E_{\Gamma_e}^{(0)}=-\nabla\varphi^{(0)}\vert_{\Gamma_e}\cdot\vec{n}$. By letting the solver start below the onset voltage, this initialization corresponds to the exact solution to System~\ref{eq: DD}.
Algorithm~\ref{alg: Corona} shows in detail the corona solver strategy.
\begin{algorithm}[H]
\caption{Corona solver}
\setlength{\intextsep}{3pt}    
\setlength{\textfloatsep}{3pt} 
\setlength{\floatsep}{3pt}     
\begin{algorithmic}[1]
    \STATE {Initial guesses}
    \STATE $n_p^{(0)} \gets n_{min}$, \; $n_n^{(0)} \gets n_{min}$ in $\Omega$
    \STATE Compute $\varphi^{(0)}$ by solving the Poisson equation with $\varphi = \pm V_0$ on $\Gamma_e$ and $\varphi = 0$ on $\Gamma_c$
    \STATE $E_{\Gamma_e}^{(0)} \gets -\nabla \varphi^{(0)} \cdot \hat{n}$ on $\Gamma_e$
    \STATE $\mathbf{z}_0^{(0)} \gets \big(\varphi^{(0)}, n_p^{(0)}, n_n^{(0)}, E_{\Gamma_e}^{(0)}\big)$
    \vspace{0.01em}
    \WHILE{$V_j < V_{final}$}
    \vspace{0.01em}
    \vspace{0.01em}
    \STATE{Set boundary conditions on $z_j^{(0)}$}
    \vspace{0.01em}  
    \WHILE{not converged and $k < k_{max}$}
        \vspace{0.01em}
        \STATE{Assemble the residuals $\mathcal{R}(z_j^{(k)})$, and the Jacobian matrix $\mathbb{J}(z_j^{(k)})$}
        \vspace{0.01em}
        \STATE{Solve the linear system: $\mathbb{J}(z_j^{(k)}) \, \delta z_j^{(k)} = -\mathcal{R}(z_j^{(k)})$ with boundary conditions on $\delta z_j^{(k)}$}
        \vspace{0.01em}
        \STATE{Update the solution: $z_j^{(k+1)} \gets z_j^{(k)} + \delta z_j^{(k)}$}
        \vspace{0.01em}
        \STATE{$k \gets k + 1$}
        \vspace{0.01em}
    \ENDWHILE
    \vspace{0.01em}
    \STATE{Set: $z_{j+1}^{(0)} \gets z_j^{(k)}$}
    \STATE{Set: $V_{j+1} \gets V_j$}
    \vspace{0.01em}
    \STATE{$j \gets j + 1$}
    \vspace{0.01em}
    \ENDWHILE
    \vspace{0.01em}
   \end{algorithmic}
   \label{alg: Corona}
\end{algorithm}

\subsection{Numerical Solution of the Fluid Problem}

The spatial discretization of the incompressible Navier-Stokes equations is performed with the classical Galerkin Finite Element Method (FEM)~\cite{John2016}. Additional details on well-posedness of the formulation with regularized directional-do-nothing boundary condition are available in Ref.~\cite{Nogueira2025}.\\

In this section, referring to Fig.~\ref{fig: domain}, we denote with $\Gamma_N= \Gamma_{top}\,\cup\,\Gamma_{bot}$ the part of the boundary where the free-slip boundary conditions are applied; similarly, $\Gamma_D= \Gamma_{e}\,\cup\,\Gamma_{c}$ denotes the part of the boundary where no-slip boundary conditions are applied; finally, $\Gamma_{RDDN}= \Gamma_{in}\,\cup\,\Gamma_{out}$ is the part of the boundary where the RDDN condition is applied.

Considering the functional spaces:
\begin{align*}
Q &= L^2(\Omega) = \left\{q : \Omega \rightarrow \mathbb{R} \;\middle|\; \int_\Omega q^2 \, d\Omega < +\infty \right\} \\
V &= [H^1(\Omega)]^d = \left\{ \vec{v} : \Omega \rightarrow \mathbb{R}^d \;\middle|\; \int_\Omega \left( |\vec{v}|^2 + |\nabla \vec{v}|^2 \right) d\Omega < +\infty \right\} \\
V_0 &= [H^1_{D}(\Omega)]^d = \left\{ \vec{v} : \Omega \rightarrow \mathbb{R}^d \;\middle|\; \vec{v} \in V,\; \vec{v} = 0 \text{ on } \Gamma_D, \ \vec{v}\cdot\hat{n} = 0 \text{ on } \Gamma_N \right\}
\end{align*}
Let us introduce two finite dimensional spaces $V_h \subset V_0$ and $Q_h \subset Q$. 
We introduce a (vector) basis $\{\boldsymbol{\phi}_j\}_{j=1}^{\mathcal{N}_u}$ for the space $V_h$ and a (scalar) basis $\{\psi_k\}_{k=1}^{\mathcal{N}_p}$ for the space $Q_h$, where $\mathcal{N}_u = \dim(V_h)$ and $\mathcal{N}_p = \dim(Q_h)$ are the dimensions of the finite element spaces of velocity and pressure, respectively.
In this work, to avoid the need for additional stabilization due to the inf-sup condition~\cite{inf_sup_Quarteroni}, we consider the classical Taylor-Hood finite element pair in which $V_h$ is the space of vector-valued piecewise quadratic functions and $Q_h$ is the space of piecewise linear functions. Using the Galerkin approximation, we look for a discrete solution $(\vec{u}_h, p_h) \in V_h \times Q_h$ such that:
\begin{equation}
\left\{
\begin{aligned}
&\int_{\Omega}\frac{\partial \vec{u}_h}{\partial t} \cdot \vec{v}_h \ d\Omega + a(\vec{u}_h,\vec{v}_h) + c(\vec{u}_h,\vec{u}_h,\vec{v}_h)+b(\vec{u}_h, p_h)= L(\vec{v}_h) \quad \forall \ \vec{v}_h\in V_h\\
& \ b(\vec{u}_h, q_h)=0 \quad  \forall \ q_h \in Q_h
\end{aligned} 
\right. \label{eq: ns_discrete}
\end{equation}
where we introduced
the linear functional
     $L(\vec{v})=\int_{\Omega}\vec{f}\cdot \vec{v} d\Omega$; 
the bilinear forms
$a(\vec{u},\vec{v}) = \int_{\Omega} \nu \nabla \vec{u}:\nabla \vec{v} \, d\Omega$ and
$b(\vec{v},p) = -\int_{\Omega} (\nabla \cdot \vec{v})\, p \, d\Omega$; and the trilinear form
$c(\vec{u},\vec{w},\vec{v})=  \int_{\Omega}[(\vec{u}\cdot\nabla)\vec{w}]\cdot\vec{v}\ d\Omega \ + \ \frac{1}{2} \int_{\Gamma_{RDDN}} [\vec{u} \cdot \hat{n}]_\delta (\vec{v} \cdot \vec{w}) \ d\Gamma$.\\
To integrate the space discretized problem~\eqref{eq: ns_discrete} in time, we employ a composite strategy. In a first stage we make use of a first order semi-implicit Euler method.
The nonlinear term $\left( \vec{u}_h^{~n+1} \cdot \nabla \right) \vec{u}_h^{~n+1}$, that appears in the three-term form $c(\vec{u}_h,\vec{u}_h,\vec{v}_h)$, is approximated at time step $t^{n+1}$ as: $\left( \vec{u}_h^{~n} \cdot \nabla \right) \vec{u}_h^{~n+1}$. This choice allows one to avoid the non-linearity of the Navier-Stokes system of equations, while preserving a satisfactory level of accuracy. Furthermore, it does not strictly require the satisfaction of a CFL condition~\cite{cfl_Quarteroni}, thereby providing an efficient and robust strategy to initialize the flow. After a few steps with the semi-implicit Euler scheme, we switch to the Implicit Differential Algebraic (IDA) solver from the SUNDIALS library \cite{gardner2022sundials,hindmarsh2005sundials}, for which deal.II provides an interface. IDA employs an efficient  variable order, variable-coefficient Backward Differentiation Formula (BDF) for time stepping.

\subsection{One-way Coupling Algorithm between Corona and Fluid Problems}

In this contribution we take into account a one-way coupling between the corona discharge and the fluid. The corona discharge influences the fluid with the body force $\vec{f}_{EHD}=q(n_p-n_n) \vec{E}$ in the momentum equation, but the fluid convective velocity does not affect the transport of ionized species.
This coupling strategy is justified by the fact that the effect of the fluid flow on the corona discharge is negligible when the flow is not subject to strong adverse wind~\cite{Picella,Two-way-coupling}, as in the test cases presented in this contribution. 
Fluid simulations are advanced in time until the x-direction component $T_{EAD}$ of the effective thrust reaches the steady state within an user prescribed tolerance. To assess convergence to stationary solution, the IDA solver monitors two quantities every $\hat{t}$ seconds.
The first quantity is the relative $L_2$-norm of the time derivative of the fluid velocity:
 $$
 \frac{\vert\vert\dot{\vec{u}}\vert\vert_{L_2}}{\vert\vert \vec{u} \vert\vert_{L_2}}
 $$
while the second is the relative variation of the thrust in the x-direction:
 $$
\frac{\vert T_{EAD}-T_{EAD,old}\vert}{\vert T_{EAD,old}\vert}
 $$
where $T_{EAD}$ denotes the thrust at time $t$ and $T_{EAD,old}$ at time $t-\hat{t}$. Convergence to steady state happens when one of these quantities is less than $tol$, which has been set for all the coupled simulations to $1\times10^{-7}$.
The coupled algorithm is described in Algorithm~\ref{alg: coupled_alg}.
\begin{algorithm}[H]
\caption{Coupled solver}
\begin{algorithmic}[1]
    \STATE {Solve the Corona problem as stated in algorithm \ref{alg: Corona}}
    \vspace{0.01em}
    \STATE{Compute $\vec{f}_{EHD}$ and initialize $\vec{u}_0=\mathbf{0}$}
    \STATE{Compute a consistent initial state for IDA, using semi-implicit Euler scheme}
    \STATE Set $t = 0$, $T_{EAD,old} = T_{EAD}^0$.

    \STATE \textbf{Time integration loop:}
    \WHILE{convergence criterion not satisfied}

    \STATE Advance the coupled fluid system by one IDA time step to obtain $\vec{u}^{n+1}$.
    
    \STATE Update time: $t \leftarrow t + \Delta t$.
    
    \STATE Compute effective thrust $T_{EAD}^{n+1}$.
    
    \IF{$t\mod\hat{t} = 0$}
        \STATE Compute monitoring quantities:
        \STATE \hspace{1em} $r_u = \dfrac{\lVert \dot{\vec{u}} \rVert_{L_2}}{\lVert \vec{u} \rVert_{L_2}}$
        \STATE \hspace{1em} $r_T = \dfrac{|T_{EAD}^{n+1} - T_{EAD,old}|}{\vert T_{EAD,old}\vert}$
        
        \STATE Update $T_{EAD,old} \leftarrow T_{EAD}^{n+1}$.
        
        \IF{$r_u < tol$ \textbf{or} $r_T < tol$}
            \STATE \textbf{break}
        \ENDIF
    \ENDIF

\ENDWHILE

\STATE \textbf{End simulation.}
\end{algorithmic}
\label{alg: coupled_alg}
\end{algorithm} 

\section{Positive and Negative Corona: Unipolar Validation}\label{sec: Validation}

This section presents results obtained for single-stage thruster configurations. The setup and parameters used in the simulations are taken from the experiments carried out in Ref.~\cite{GomezVega2024}. 
The experimental configuration consists of a variable number of parallel emitter-collector pairs, where each emitter is a wire of radius $R_e$ and each collector is an HT-14 profile airfoil. The gap between emitter and collector is $d$ and the distance between emitter-collector pairs is $\Delta$.
Temperature of ions and neutrals, density and viscosity of the fluid, mobility of positive and negative ions are
given in Table~\ref{tab: physical_quantitites}. The values of mobility of positive and negative ions are taken from Ref.~\cite{GomezVega2024}.
\begin{table}[H]
\centering
\begin{tabular}{l l l l}
  \toprule
  \textbf{Symbol} & & \textbf{Value}  & \textbf{Units} \\
    $T$  & & $300$ & $[\,\text{K}\,]$  \\
    $\rho_f$ & &  $1.225$ &  $[\,\text{kg}\,\text{m}^{-3}\,]$  \\
    $\mu_f$ & & $1.8\times10^{-5}$ & $[\,\text{kg}\,\text{m}^{-1}\,\text{s}^{-1}\,]$ \\
    $\mu_p$  & &  $2\times10^{-4}$ &  $[\,\text{m}^{2}\,\text{V}^{-1}\,\text{s}^{-1}\,]$   \\
    $\mu_n$  & &  $2.3\times10^{-4}$ &  $[\,\text{m}^{2}\,\text{V}^{-1}\,\text{s}^{-1}\,]$
\end{tabular}
\caption{Physical Quantities}\label{tab: physical_quantitites}
\end{table}
The other parameters used in the simulations are summarized in Table~\ref{tab: single_stage_parameters}. We specify the geometric characteristics of the electrodes, the applied voltage and the parameters for the injection law given by Eqs.~\ref{eq: BC_ion}. The gap $d$ and the distance between pairs $\Delta$ vary between different simulations.
\begin{table}[H]
\centering
\begin{tabular}{l l l l l}
  \toprule
  \textbf{Variable} & \textbf{} & \textbf{Value} & &\textbf{Units}  \\
    \text{$R_{e}$} &  & $28$                     & & $[\,\mu \text{m}\,]$     \\
    \text{HT-14 chord} &  & $20$                 & & $[\,\text{mm}\,]$         \\
    \text{HT-14 thickness}  & & $2$              & & $[\,\text{mm}\,]$       \\
    \text{$V_a$}    & & $1$--$15$                & & $[\,\text{kV}\,]$         \\
    \text{$n_{min}$} & & $10^{9}$                 & & $[\,\text{m}^{-3}\,]$    \\
    \text{$n_{ref}$} & & $10^{9}$                 & & $[\,\text{m}^{-3}\,]$    \\
    \text{$E_{ref}$} & & $2.3917\times 10^{5}$    & & $[\,\text{V}\,\text{m}^{-1}\,]$      \\
    \text{$\Bar{E}_{on}$} & & $1.7938\times10^{7}$     & & $[\,\text{V}\,\text{m}^{-1}\,]$   \\
\end{tabular}
\caption{Parameters for the setup of Ref.~\cite{GomezVega2024}.}
\label{tab: single_stage_parameters}
\end{table}

\subsection{Grid Convergence Analysis for the Corona Problem} \label{subsec: single_stage_grid_convergence}

A grid convergence analysis was performed to test the accuracy of the numerical results. We show the convergence behavior in the case of a single emitter-collector pair with a gap $d=20\,\text{mm}$ and a positive corona discharge. 
We use the same mesh for the corona and the fluid problem. The quantities monitored for convergence are the current $I$ and the horizontal component of the electrostatic thrust $\vec{T}_{E}$ for the corona problem and the horizontal component of the electro-aerodynamic thrust $\vec{T}_{EAD}$ for the fluid problem.

\begin{figure}[H]
    \centering
     \includegraphics[width=0.48\textwidth]{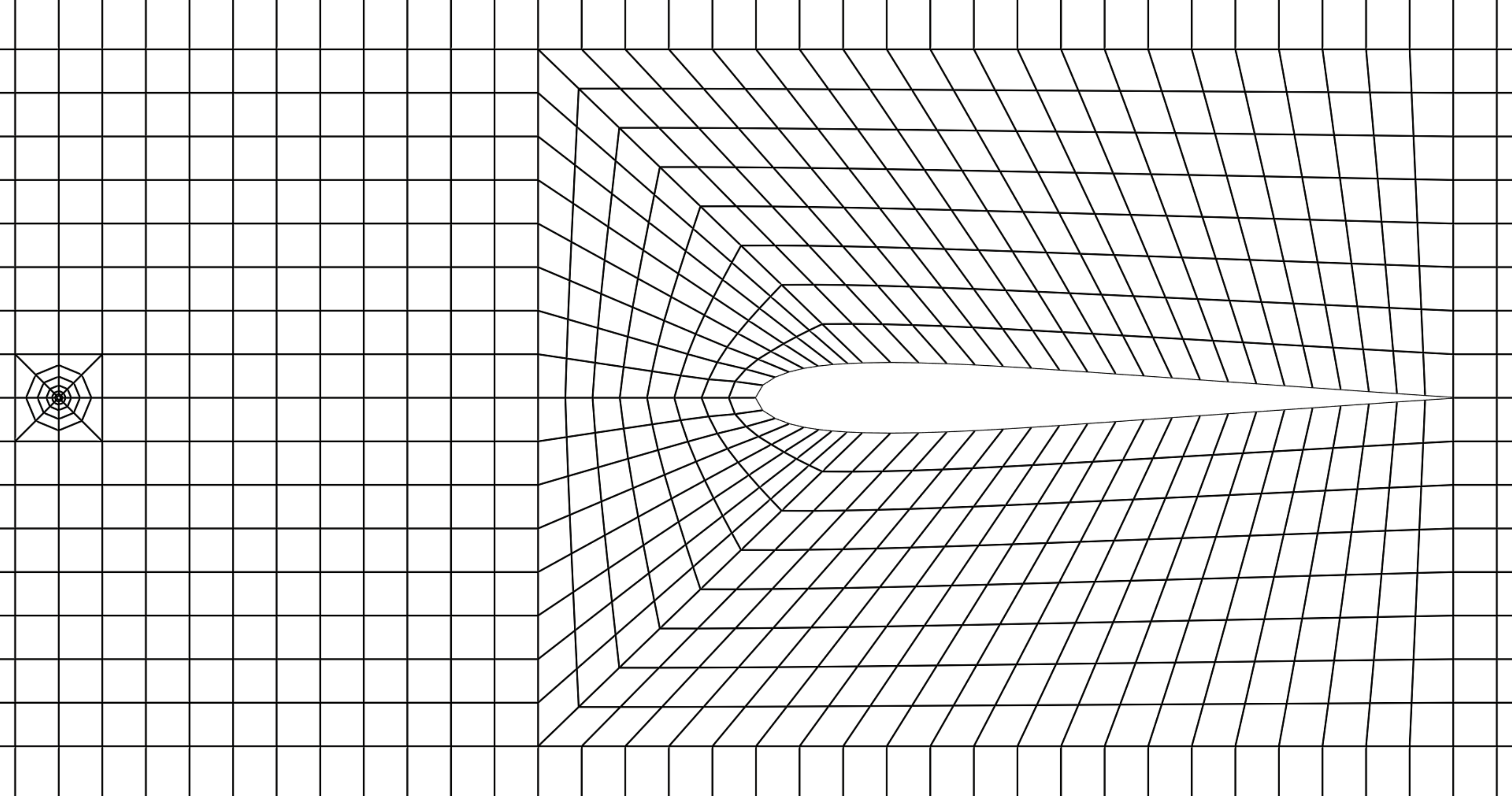}
     \caption{Zoom near an emitter-collector pair of the base coarse mesh. Single emitter-collector pair, gap $d=20\, \text{mm}$.}
     \label{fig: single_stage_meshes_coarse}
\end{figure}

\begin{figure}[H]
    \centering
    \includegraphics[width=0.48\textwidth]{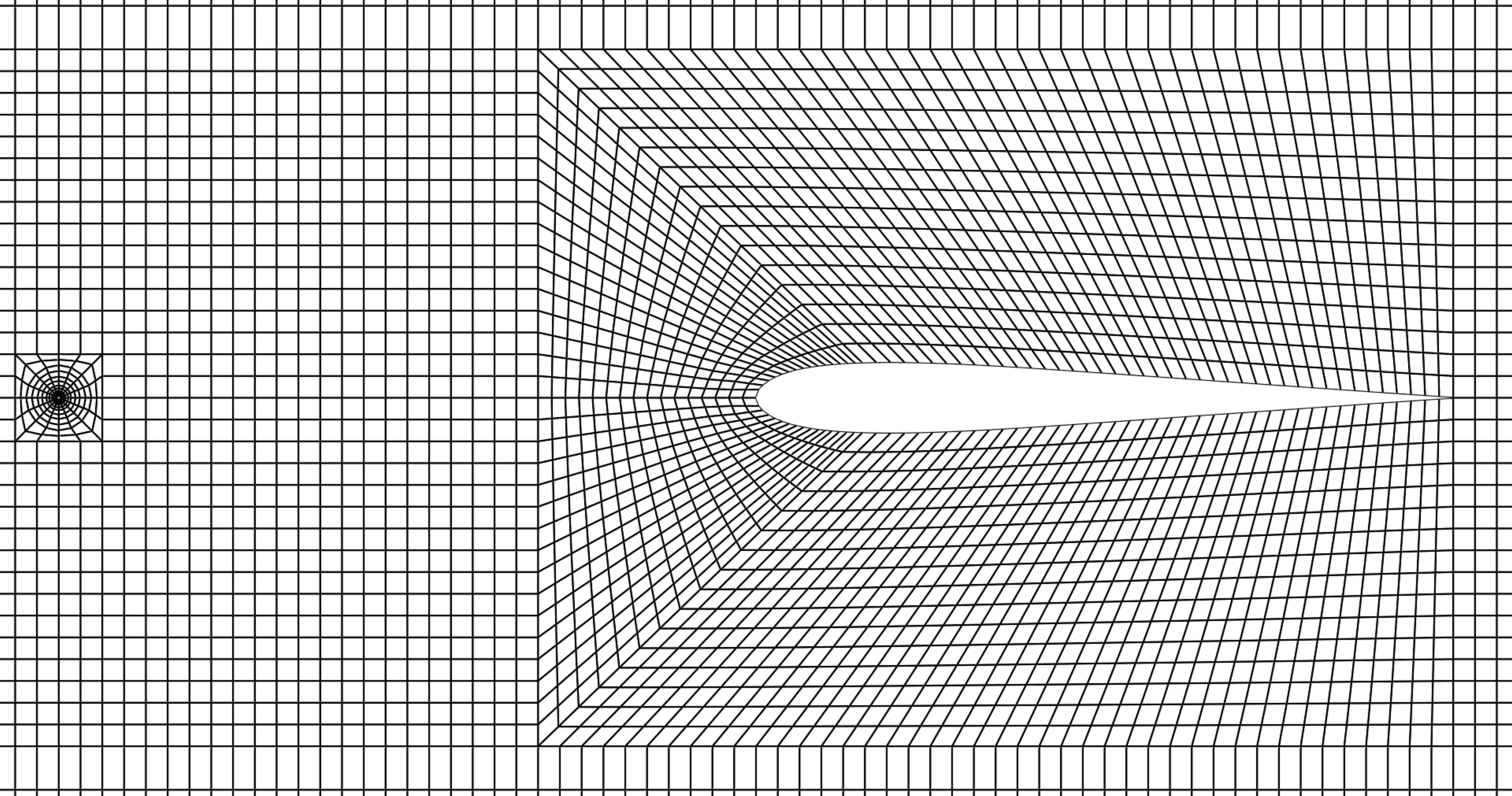}
    \caption[Meshes used in the grid convergence analysis]{Zoom near an emitter-collector pair of the one-step globally refined mesh. Single emitter-collector pair, gap $d=20\, \text{mm}$.}
    \label{fig: single_stage_meshes_global1}
\end{figure}

\begin{figure}[H]
    \centering
    \includegraphics[width=0.48\textwidth]{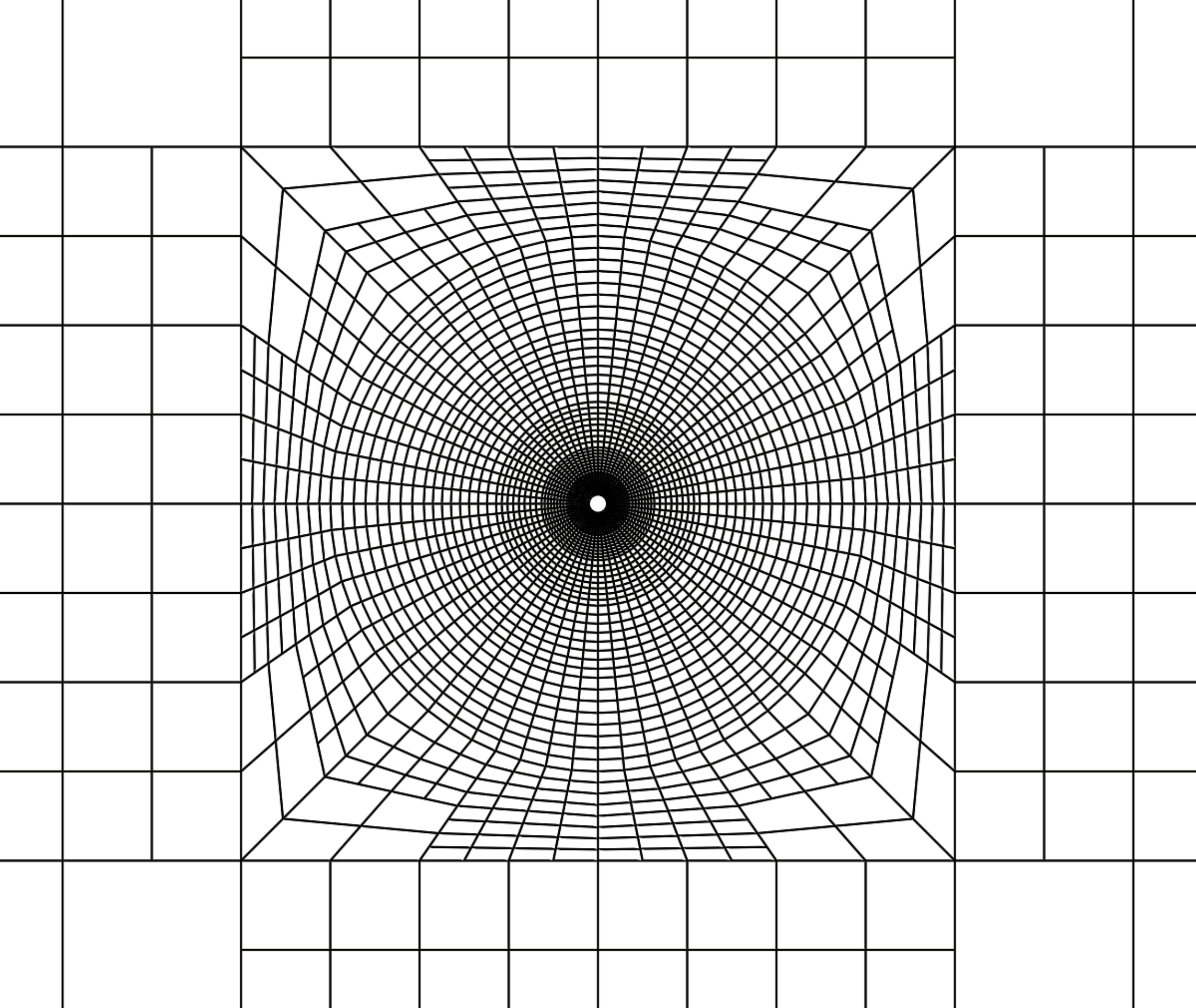}
    \caption{Zoomed-in view of local refinements around the emitter. Mesh after one global and two local refinements.}
    \label{fig: single_stage_local_refinements_emitter_g1l2}
\end{figure}

\begin{figure}[H]  
    \centering
    \includegraphics[width=0.48\textwidth]{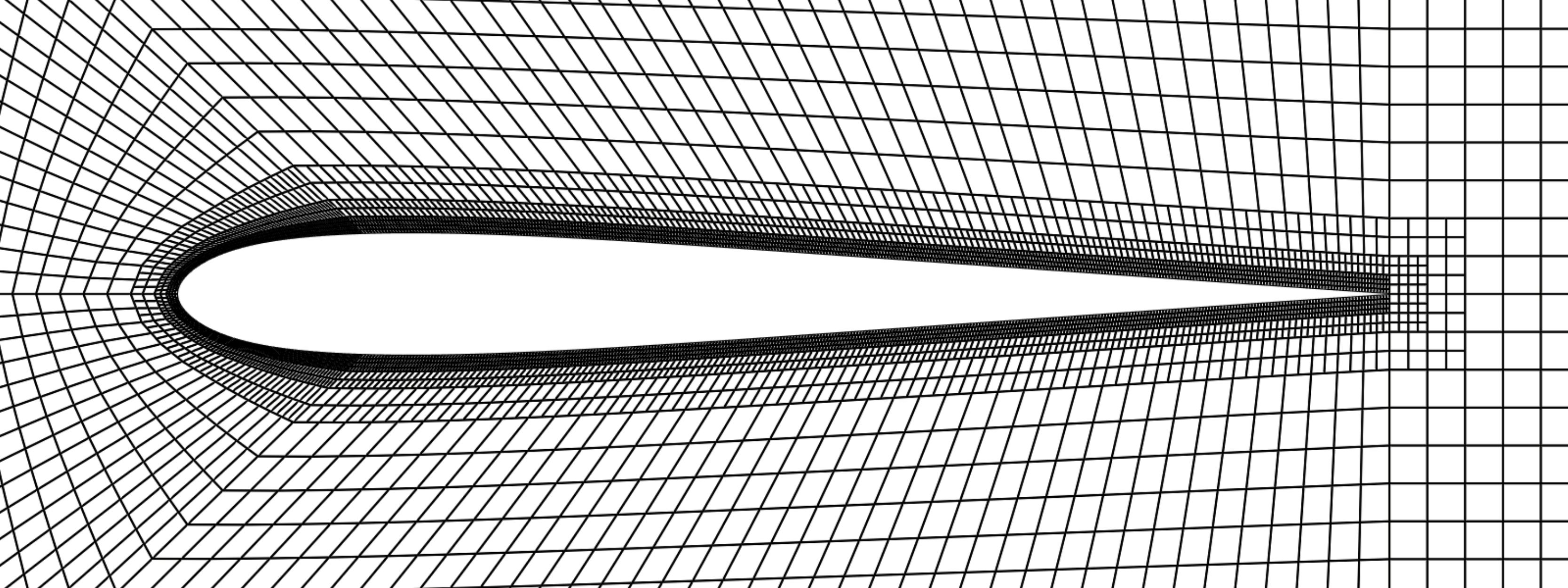}
    \caption{Zoomed-in view of the local refinements around the collector. Mesh after one global and three local refinements.}
    \label{fig: single_stage_local_refinements_collector_g1l3}
\end{figure}

The convergence analysis involved simulations with successive meshes that are obtained by refining a base coarse mesh globally and locally.
The combination of global and local refinements is used to obtain accurate results without excessive computational cost. The local refinement focuses on the regions near the emitter and collector surfaces. The number of local refinements close to the emitter can differ from the number of local refinements close to the collector.
We use six different meshes and their properties are summarized in Table~\ref{tab: single_stage_grid_convergence}. For meshes obtained with a combination of global and local refinements, global refinement is performed first, followed by local refinements.
\begin{table}[H] 
    \centering 
    \begin{tabular}{c c c c}
    \toprule
     \textbf{Mesh} & \textbf{Global refinements} & \textbf{Emitter refinements} & \textbf{Collector refinements} \\
    1   & 0              & 0           & 0        \\[1pt]
    2   & 0              & 1           & 1        \\[1pt]
    3   & 1              & 1           & 3        \\[1pt]
    4   & 1              & 2           & 6        \\[1pt]
    5   & 1              & 3           & 7        \\[1pt]
    6   & 2              & 3           & 7        \\[1pt]
    \end{tabular}
    \caption{Refinement strategy used in the grid convergence analysis for the single-stage thruster configuration. Single emitter-collector pair, gap $d=20\, \text{mm}$. 
    }
    \label{tab: single_stage_grid_convergence}
\end{table}
Zooms of the base coarse mesh and of the mesh after one global refinement are shown in Fig.~\ref{fig: single_stage_meshes_coarse} and in Fig.~\ref{fig: single_stage_meshes_global1} respectively.
Zoomed-in views of local refinements around the emitter are shown in Fig.~\ref{fig: single_stage_local_refinements_emitter_g1l2}; and around the collector in Fig.~\ref{fig: single_stage_local_refinements_collector_g1l3}.

\begin{figure}[H]
    \centering
        \includegraphics[width=0.53\textwidth]{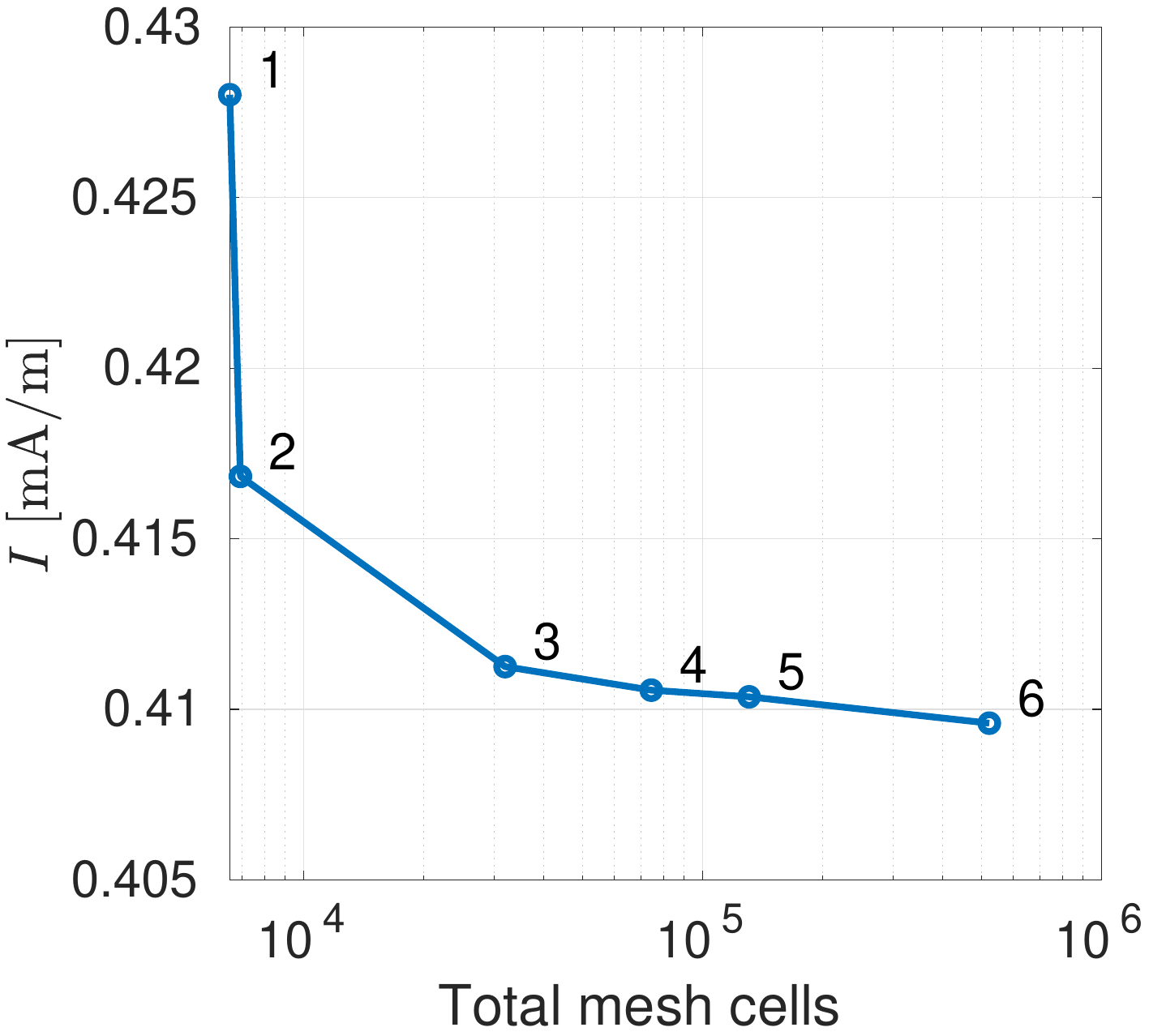}
    \caption{Grid convergence analysis. Single emitter-collector pair, gap $d=20\, \text{mm}$.  Current $I$ vs.\ number of cells.}
     \label{fig: I_single_stage_grid_convergence}
\end{figure}

\begin{figure}[H]
    \centering
        \includegraphics[width=0.53\textwidth]{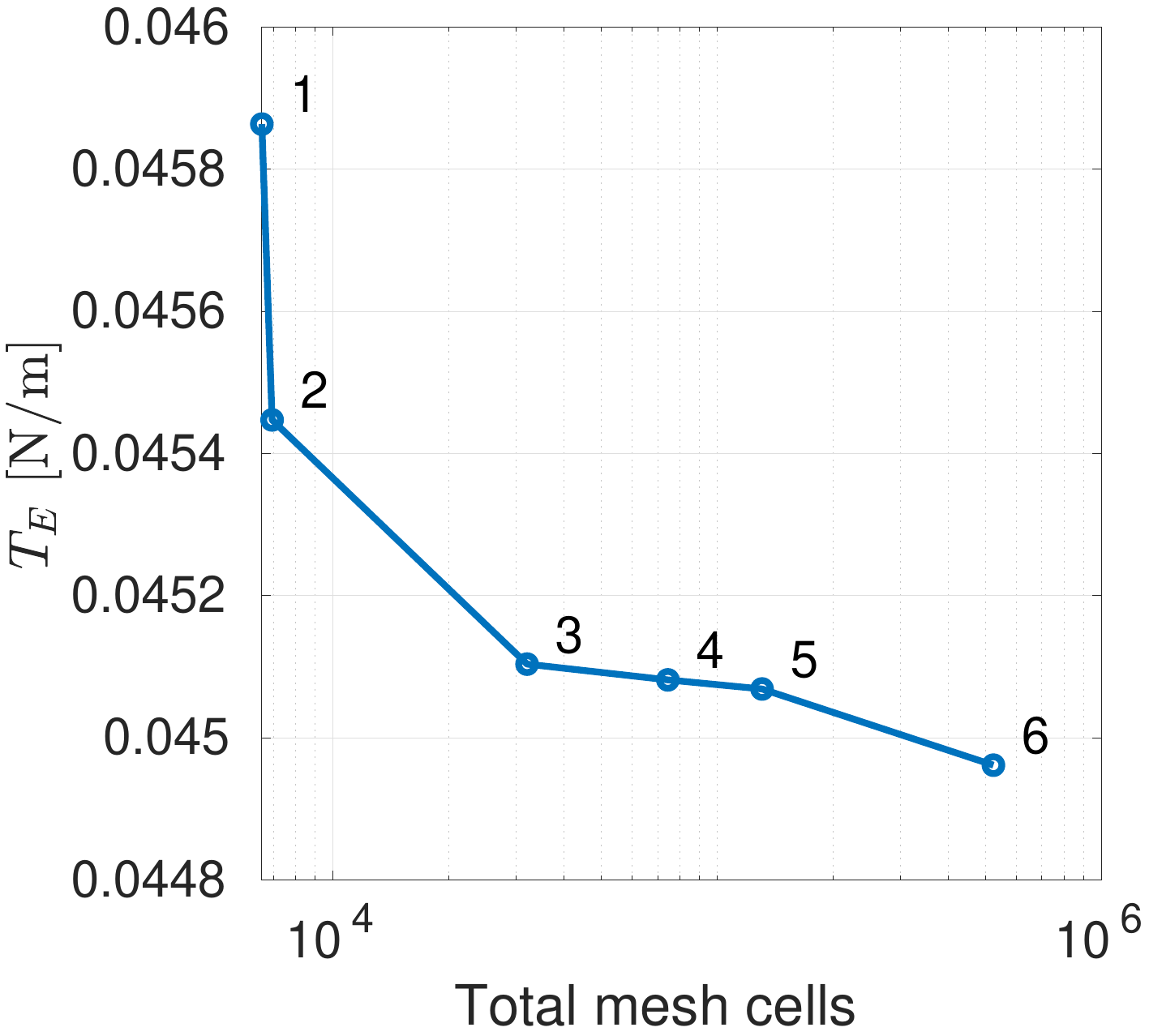}
        \caption{Grid convergence analysis. Single emitter-collector pair, gap $d=20\, \text{mm}$. Horizontal component of electrostatic thrust $\vec{T}_{E}$  vs.\ number of cells.}
    \label{fig: T_E_single_stage_grid_convergence}
\end{figure}

The results of the grid convergence analysis for the corona problem are summarized in Figs.~\ref{fig: I_single_stage_grid_convergence} and~\ref{fig: T_E_single_stage_grid_convergence}, that show the computed current $I$ and electrostatic thrust $T_{E}$ as functions of the number of cells in the mesh. The results indicate that both quantities converge with mesh refinement, with variations below $0.3\,\%$ in both the current and the thrust across the last three meshes considered.

\begin{figure}[!ht]
    \centering
    \includegraphics[width=0.53\textwidth]{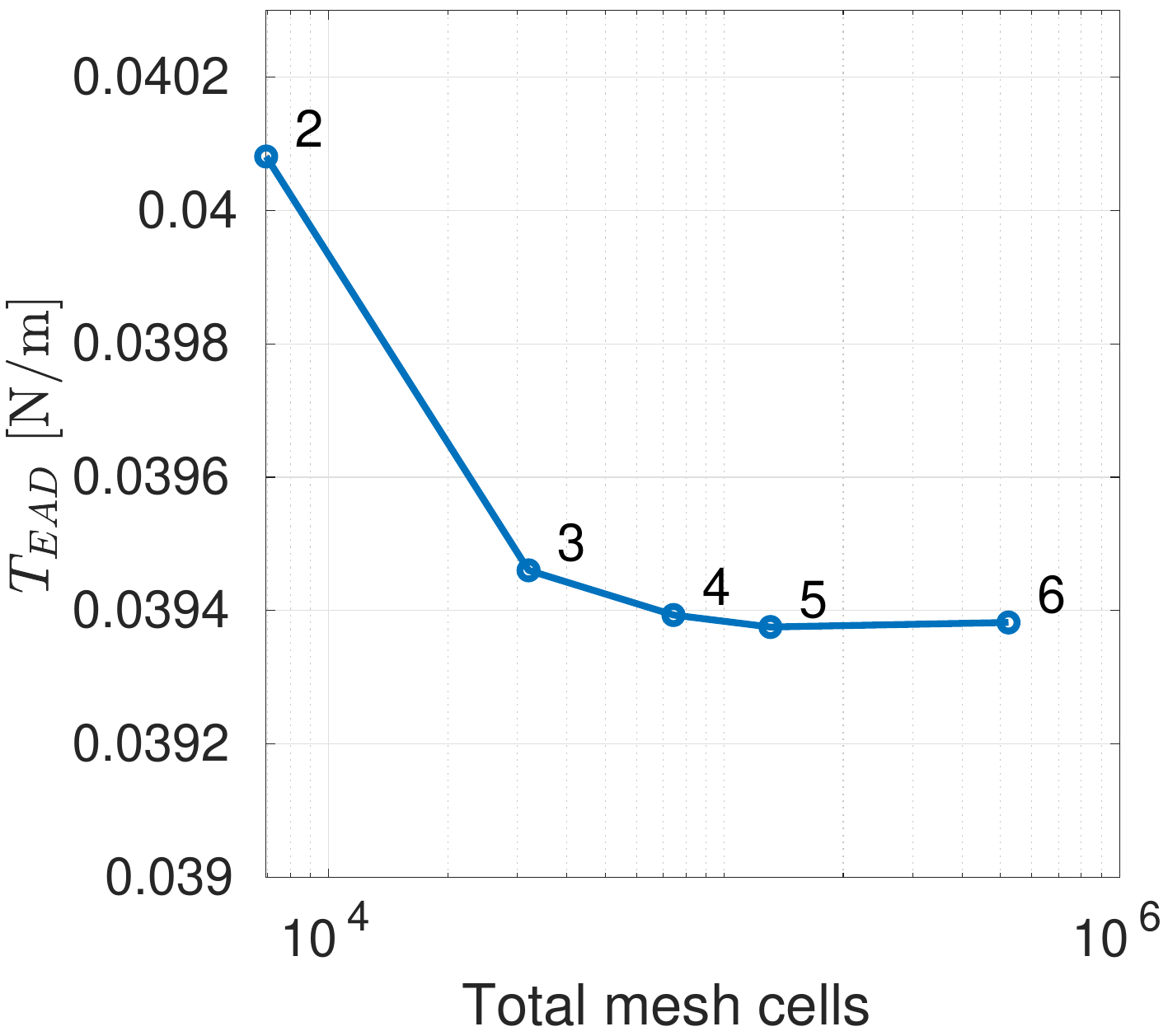}
    \caption{Grid convergence analysis. Single emitter-collector pair, gap $d=20\, \text{mm}$. Horizontal component of EAD thrust $\vec{T}_{EAD}$  vs.\ number of cells.}
    \label{fig:net_thrust_conv}
\end{figure}

The effect of grid refinement on the horizontal component of electro-aerodynamic thrust $\vec{T}_{EAD}$ is shown in Fig.~\ref{fig:net_thrust_conv}. The relative difference in thrust among the last three meshes does not exceed $0.05 \%$.

\subsection{Validation for the Corona Problem}
\label{sec:validation}
\begin{figure}[H]
    \centering
        \includegraphics[width=0.53\textwidth]{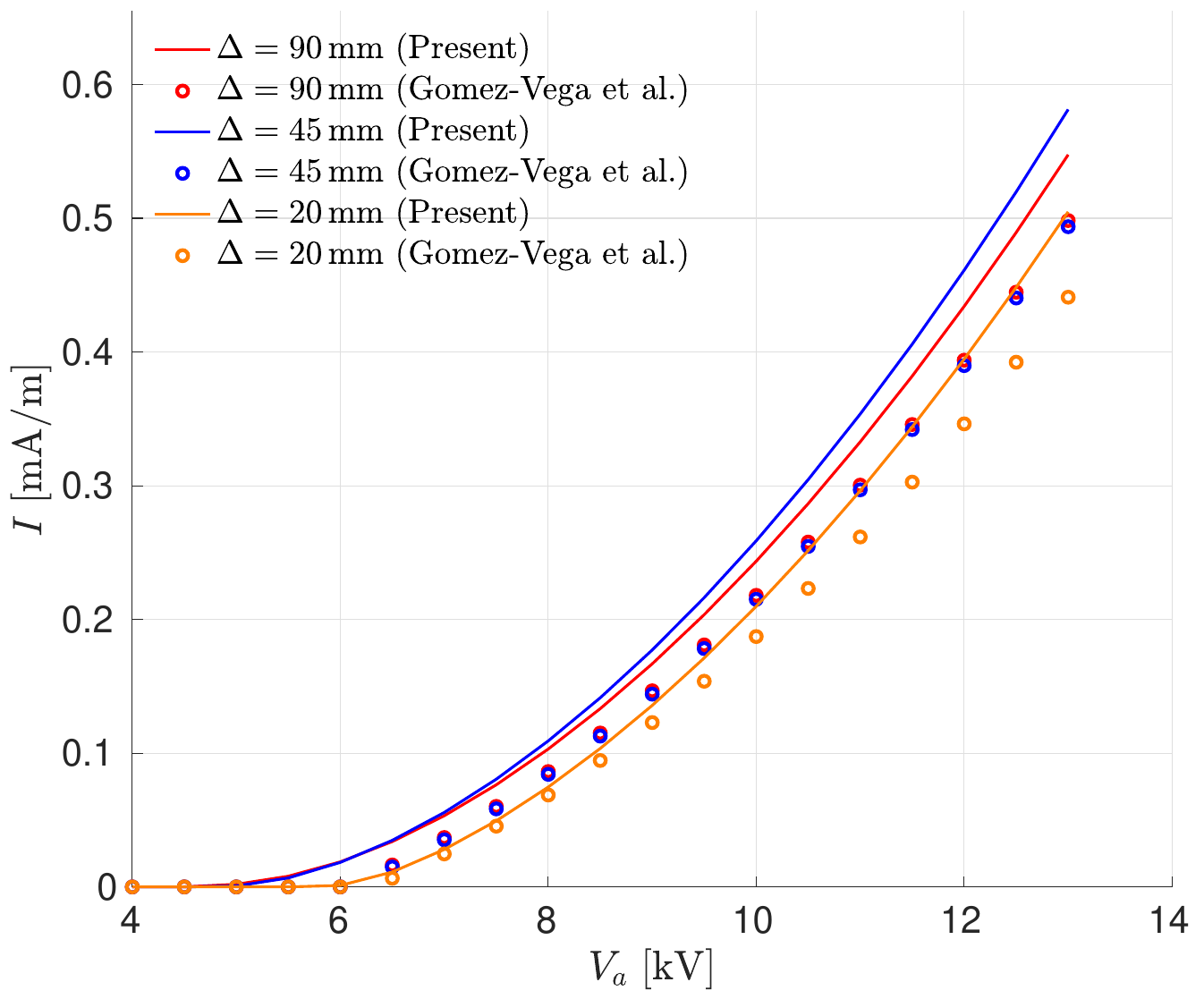}
        \caption{Positive corona. Current-voltage curves. Gap $d=15$ mm and variable spacing $\Delta$. Comparison with the model of Ref.~\cite{GomezVega2024}.}
        \label{fig: posc_single_stage_results_d15}
\end{figure}

\begin{figure}[H]
    \centering
        \includegraphics[width=0.53\textwidth]{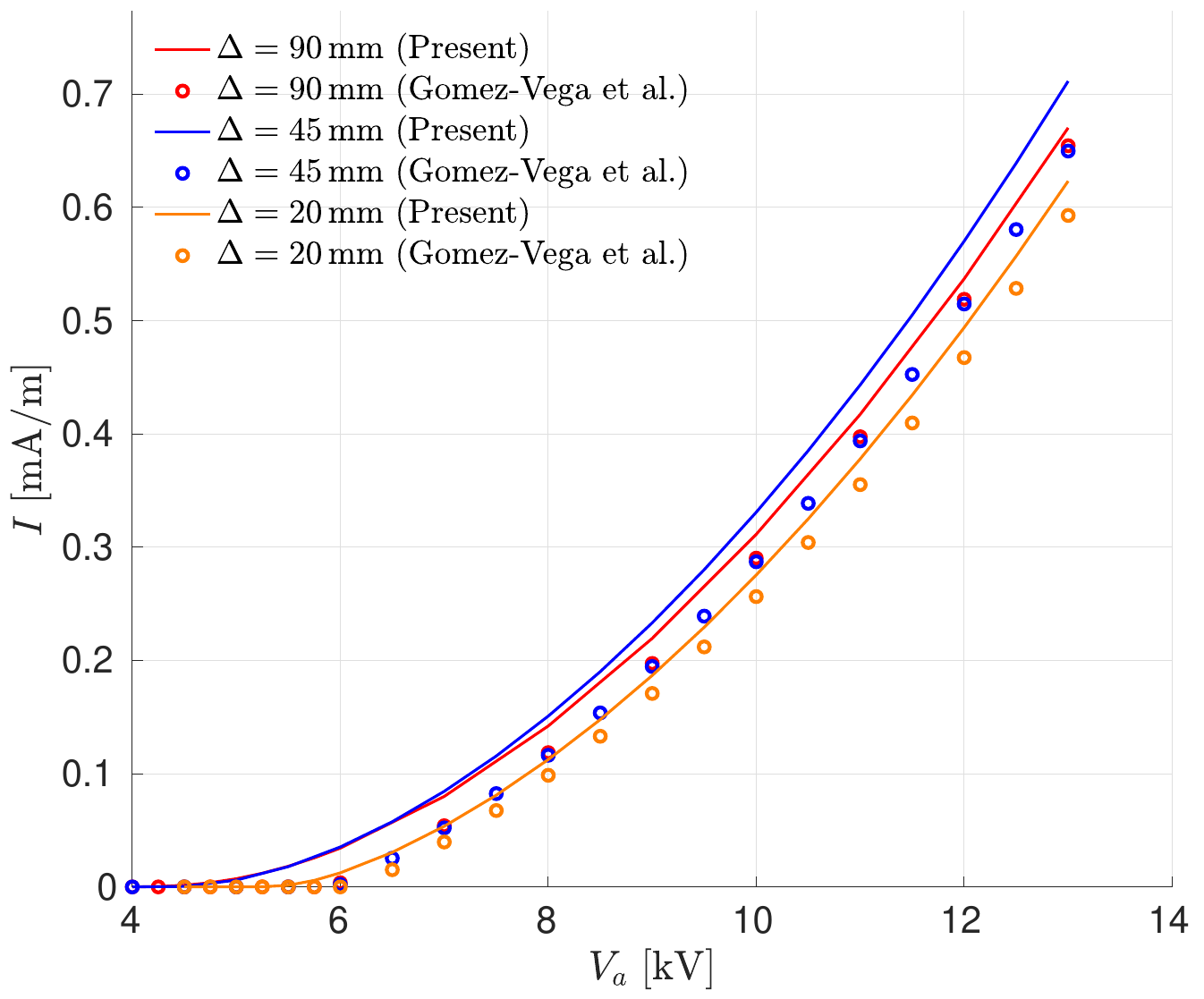}
    \caption{Negative corona. Current-voltage curves. Gap $d=15$ mm and variable spacing $\Delta$. Comparison with the model of Ref.~\cite{GomezVega2024}.}
    \label{fig: negc_single_stage_results_d15}
\end{figure}

\begin{figure}[H]
    \centering
        \includegraphics[width=0.53\textwidth]{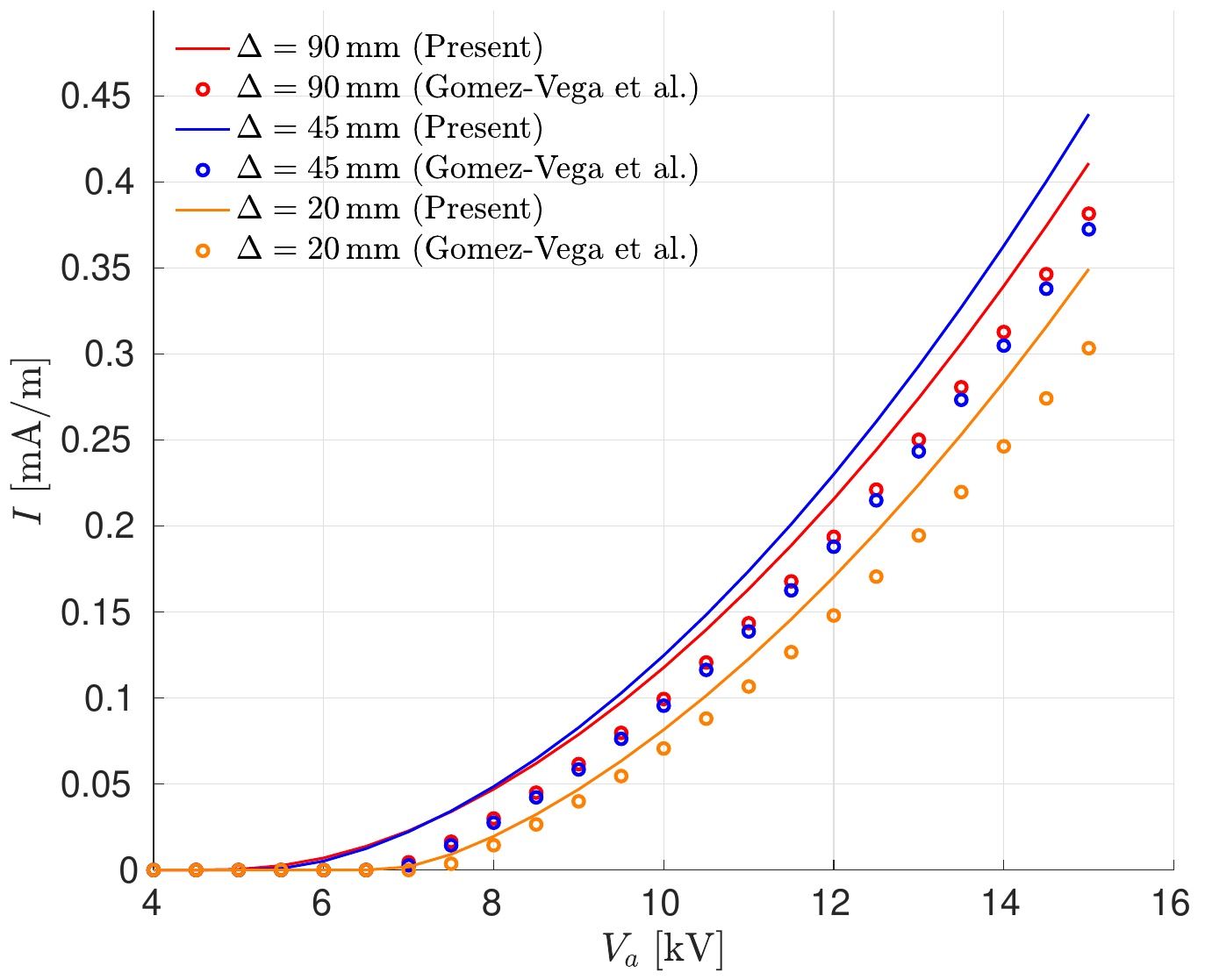}
    \caption{Positive corona. Current-voltage curves. Gap $d=20$ mm and variable spacing $\Delta$. Comparison with the model of Ref.~\cite{GomezVega2024}.}
    \label{fig: posc_single_stage_results_d20}
\end{figure}

\begin{figure}[H]
\centering
    \includegraphics[width=0.53\textwidth]{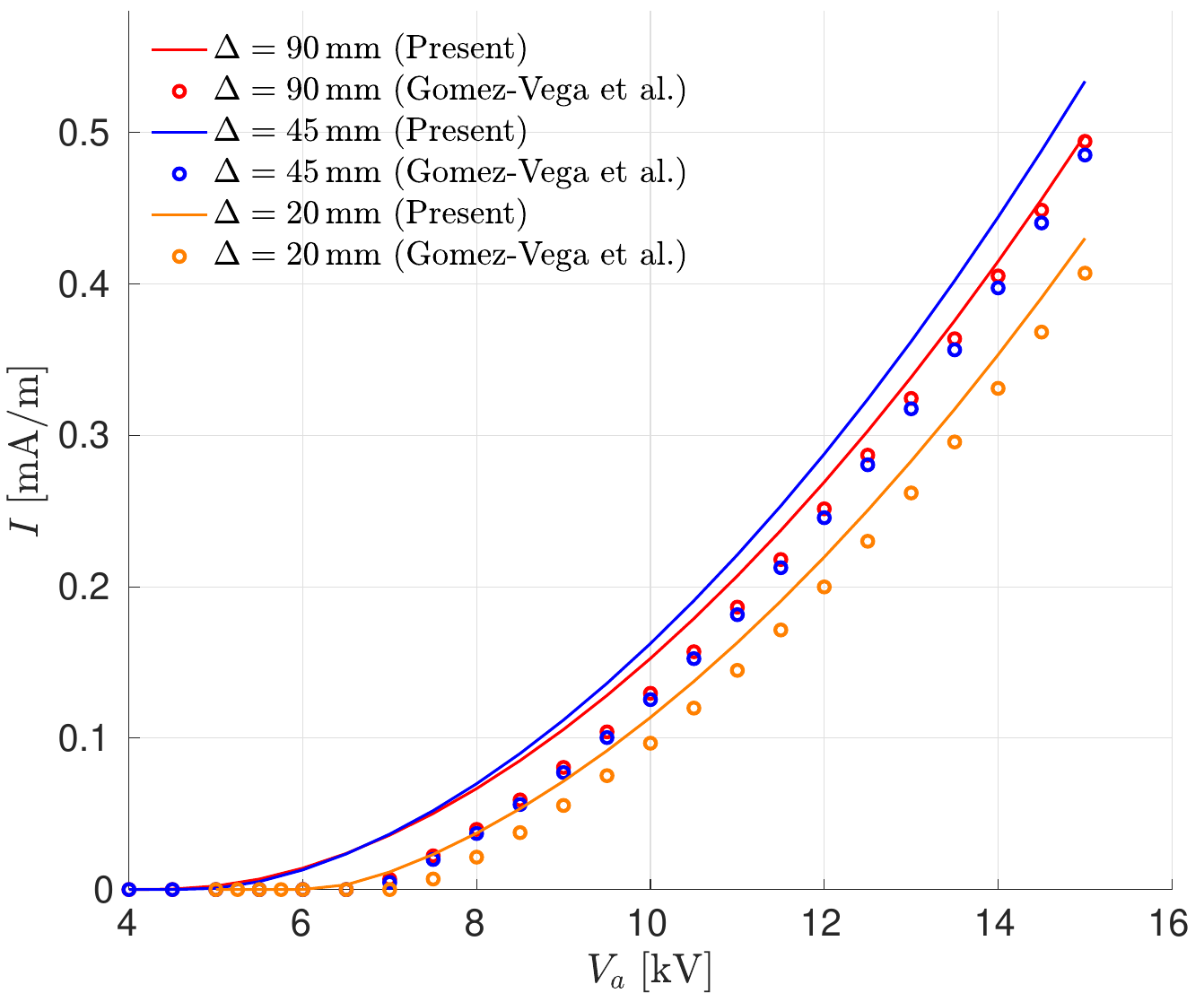}
    \caption{Negative corona. Current-voltage curves. Gap $d=20$ mm and variable spacing $\Delta$. Comparison with the model of Ref.~\cite{GomezVega2024}.}
    \label{fig: negc_single_stage_results_d20}
\end{figure}

Based on the previous grid convergence analysis, Mesh 5 of Table~\ref{tab: single_stage_grid_convergence} was selected for the simulations of the single-stage thruster configuration shown in this section. This mesh provided a good balance between accuracy and computational cost, ensuring reliable results for the analysis of the thruster performance.
We focus our attention on the performance metrics of current $I$ and electrostatic thrust $\vec{T}_{E}$ as functions of the applied voltage $V_a$.
We recall that the current $I$ per unit length is computed with Eq.~\ref{eq:current} and the thrust is computed with Eq.~\ref{eq:Tehd}.
The results are compared with the model developed in Ref.~\cite{GomezVega2024} and based on the fit of experimental data. This choice was made to allow comparisons across different geometries without the need for experimental data for each specific configuration.
In the model of Ref.~\cite{GomezVega2024} the current $I$ per unit length as a function of the voltage $V_a$ is given by the formula:
\begin{equation}
I =
\begin{cases}
0, & \text{if } |V_a| < |V_0|,\\[6pt]
C_0\left(\frac{\Delta}{d}\right) \, \varepsilon \, \mu\,
\dfrac{V_a\left[V_a - V_0\left(d,\Delta/d\right)\right]}{d^{2}}, & \text{if } |V_a| \ge |V_0|.
\end{cases}
 \label{eq:barrett_iv}
\end{equation}
where $\mu$ is the mobility of ions (different for positive and negative ions), $d$ is the gap between emitter and collector, and $C_0$, $V_0$ are empirical functions of electrodes geometry, gap, spacing between electrode pairs and corona polarity. The values of $C_0$ and $V_0$ were taken from Ref.~\cite{GomezVega2024} for each configuration considered. \\
In the numerical computations, three different spacings between emitter-collector pairs were considered: $\Delta = 90\, \text{mm}$, $\Delta = 45\,\text{mm}$ and $\Delta = 20\,\text{mm}$, that, with a total height of the simulated domain of $180\,\text{mm}$, correspond to configurations with $1$, $4$, and $9$ pairs respectively. For each spacing, simulations were performed with gaps $d = 15\,\text{mm}$ and $d = 20\,\text{mm}$ for both positive and negative corona discharges; all the simulations were carried out with the set of parameters of Tables~\ref{tab: physical_quantitites} and~\ref{tab: single_stage_parameters}. \\
The current-voltage curves are shown in Figs.~\ref{fig: posc_single_stage_results_d15} and~\ref{fig: posc_single_stage_results_d20} for positive corona discharges and in Figs.~\ref{fig: negc_single_stage_results_d15} and~\ref{fig: negc_single_stage_results_d20} for negative corona discharges.
In the plots the current $I$ is normalized by the number of emitter-collector pairs in order to obtain the current per pair. 
Equation~\ref{eq:barrett_iv} predicts that, for a given applied voltage, the current drawn by each emitter-collector pair decreases as the spacing $\Delta$ between pairs decreases. The experimental results of Ref.~\cite{GomezVega2024} show this trend for small $\Delta/d$ ratios, while very tiny increase in the current per unit span per unit pair can be noticed between $1$ and $4$ emitter-collector pairs, in particular for gap lengths of $d=15\,\text{mm}$ and $d=20\,\text{mm}$. This effect is not present in the fitting Eq.~\ref{eq:barrett_iv}, but it is reproduced by the simulations.\\
It can be observed that our results slightly overestimate the current predicted by the experimental model. The error remains within acceptable limits, with an average deviation of approximately $10\,\%$ among all the configurations considered. 


\begin{figure}[!ht]
    \centering
        \includegraphics[width=0.53\textwidth]{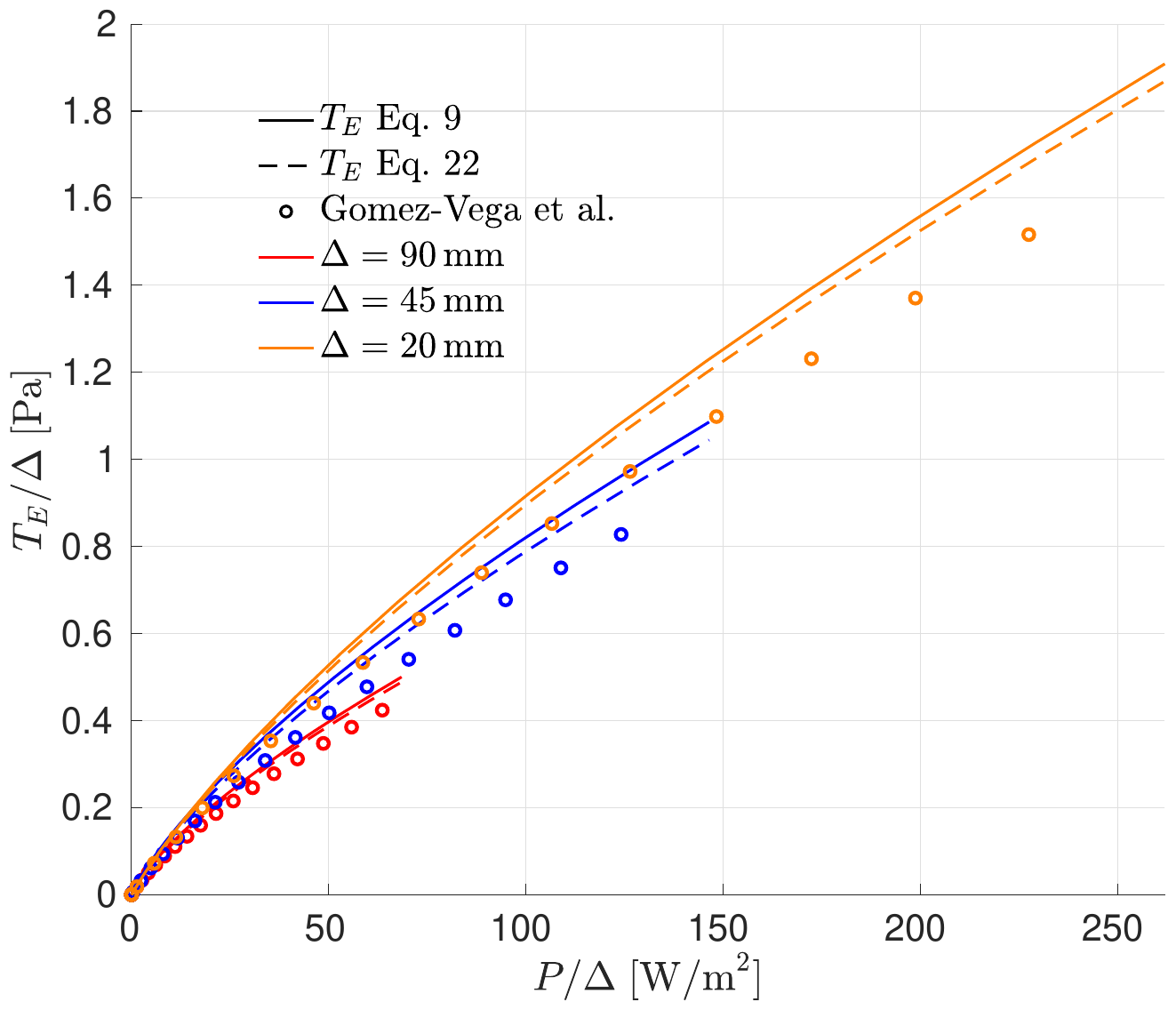}
    \caption{Thrust $T_{E}/\Delta$ for $d = 20\,\text{mm}$ in positive corona.}
        \label{fig: single_stage_thrust_results_d15}
\end{figure}

\begin{figure}[!ht]
\centering
\includegraphics[width=0.53\textwidth]{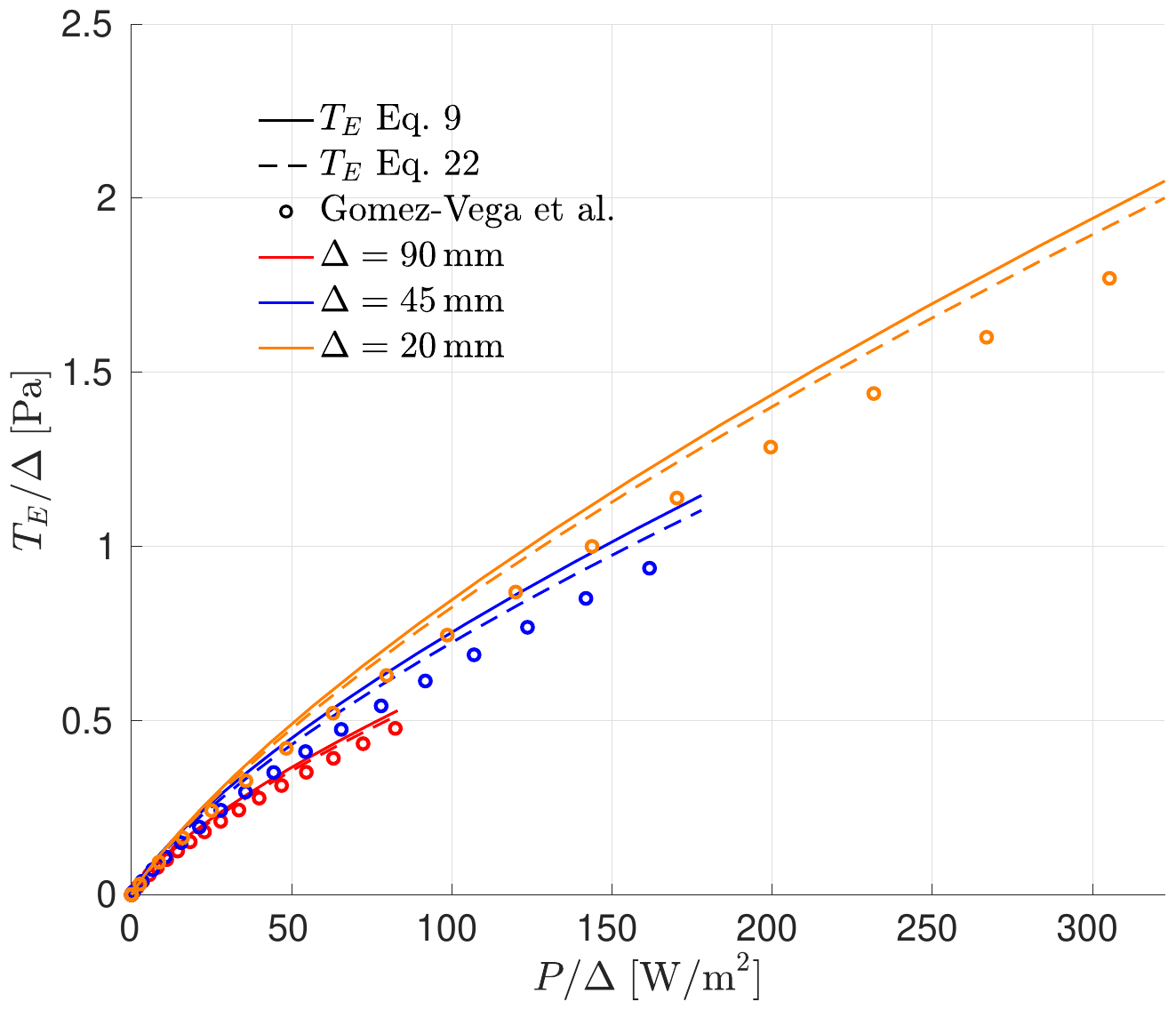}
\caption{Thrust $T_{E}/\Delta$ for $d = 20\,\text{mm}$ in negative corona}
\label{fig: single_stage_thrust_results_d20}
\end{figure}

In Figures~\ref{fig: single_stage_thrust_results_d15} and~\ref{fig: single_stage_thrust_results_d20} we show the electrostatic force per pair (computed with Eq.~\ref{eq:Tehd}) as a function of the electric power per pair for positive and negative corona respectively. Both quantities are scaled with the distance $\Delta$ between emitter-collector pairs. The thrust $T_E/\Delta$ (which is a force per unit area) corresponds to the pressure increase across an emitter-collector pair without fluid acceleration. Experimentally, it has been observed that the functional dependence of $T_{E}/\Delta$ versus $P/\Delta$ is only weakly affected by the gap $d$~\cite{GomezVega2024}; the same trend is also observed numerically. Therefore the computed electrostatic thrust values are shown only for the case $d=20\, \text{mm}$.
We compare our numerical computations with the results of Gomez-Vega et al.~\cite{GomezVega2024}. Their results are not direct experimental data, but are instead computed from the measured current with the following formula~\cite{Barrett13}:
\begin{equation}
    T_{E} = I \, d/\mu
    \label{eq:Barret_thrust}
\end{equation}
We remark that our computations predict, for the same current, a higher electrostatic thrust compared to the classical expression given by Eq.~\ref{eq:Barret_thrust}. A new scaling law, recently proposed in Ref.~\cite{Usuelli25}, can explain the underestimation of thrust associated with Eq.~\ref{eq:Barret_thrust}.
We start by defining a dimensionless current coefficient $C_C$~\cite{Usuelli25}:
\begin{equation}
    C_C=\frac{8}{9}\frac{I}{\mu\varepsilon}\frac{d^3}{V_a\Delta V}\label{eq:CC}
\end{equation}
where $\Delta V = V_a-V_{ign}$ is difference between the applied and the ignition voltage.\\
The electrostatic thrust per unit area reads~\cite{Usuelli25}:
\begin{equation}
    \frac{T_E}{A}=\varepsilon\frac{V_a\Delta V}{\chi}\frac{p}{d}C_{T_A}
    \label{eq:CTA}
\end{equation}
where $p$ is the operative pressure; $\chi=(pd)_{\text{Stoletov}}=253.15\, \text{Pa}\,\text{m}$ is a coefficient computed for the Stoletov point of the Paschen curves~\cite{Usuelli25} and $C_{T_A}$ is the dimensionless x-component of the electrostatic thrust $C_{T_A}=\int_{\hat{\Omega}} \hat{f}_{E,x}d\hat{V}$. If we solve Eq.~\ref{eq:CC} as a function of $V_a\Delta V$ and stick the result into Eq.~\ref{eq:CTA} we obtain a new expression for the electrostatic thrust: 
\begin{equation}
    T_E= \frac{Id}{\mu}\times\frac{8}{9}\frac{C_{T_A}}{C_C}\frac{pd}{\chi}\label{eq:Usuelli_thrust}
\end{equation}
If we compare the last equation for electrostatic thrust with the standard expression of Eq.~\ref{eq:Barret_thrust}, we remark that the two equations differ by a correction factor.
In the present work, the ratio $pd/\chi$
is greater than one, since the operating conditions of the corona discharge lie above the Stoletov point. The ratio between the dimensionless thrust and current $(C_{T_A}/C_C)$ depends on the specific configuration and, in principle, can be computed from experimental thrust and current measurements~\cite{Usuelli25}. Nevertheless, the numerical simulations can give a reasonable estimate. Working at atmospheric pressure, with a temperature $T=300\,\text{K}$ and a gap $d=20\, \text{mm}$, it follows that $C_{T_A}/C_C\simeq 0.15$; with these values, the correction factor appearing in Eq.~\ref{eq:Usuelli_thrust} is approximately $1.07$. Under the reasonable assumption that the same value of the correction factor applies to the experimental configuration of Ref.~\cite{GomezVega2024}, it follows that the electrostatic thrust computed with Eq.~\ref{eq:Barret_thrust} and shown in Figs.~\ref{fig: single_stage_thrust_results_d15} and~\ref{fig: single_stage_thrust_results_d20} is underestimated.
This fact is further demonstrated by the dashed curves in Figs.~\ref{fig: single_stage_thrust_results_d15} and~\ref{fig: single_stage_thrust_results_d20} that show the thrust computed with Eq.~\ref{eq:Usuelli_thrust}: not only the thrust is greater than the one computed with Eq.~\ref{eq:Barret_thrust}, but it is much closer to our numerical computations.
Finally, it can be observed that the negative corona discharge has a lower efficiency compared to the positive corona discharge, as indicated by the lower thrust achieved for the same power input.

\section{Multi-stage Thruster: an Electrical Analysis} \label{sec: Multistage}

\subsection{Effect of the Recombination Term}\label{sec: Recombination}

\begin{figure}[H]
    \centering
    \includegraphics[width=0.53\textwidth]{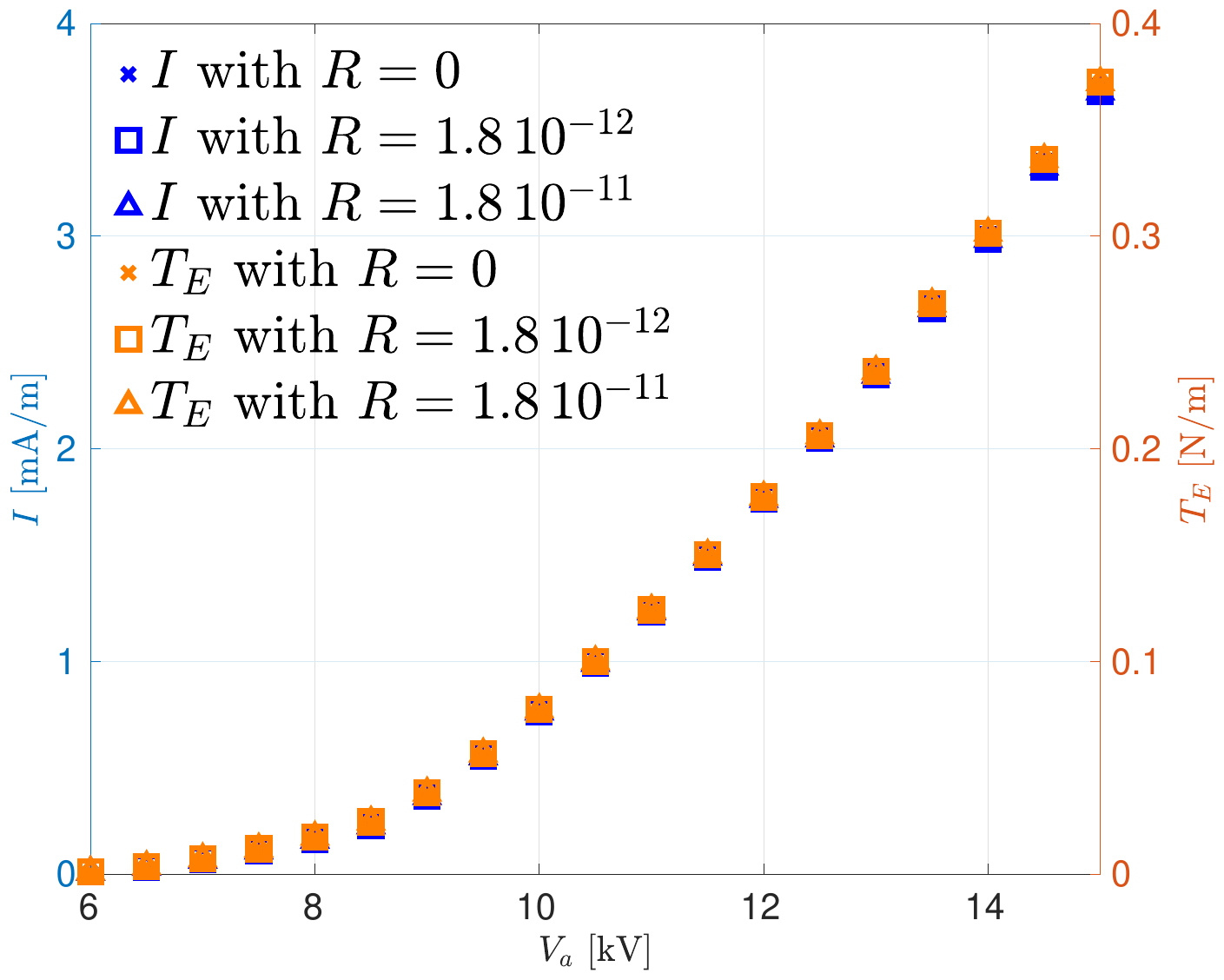}
    \caption[]{Computed current $I$ and thrust $T_{E}$ as functions of the applied voltage for three values of the recombination coefficient $R_i$.}
    \label{fig: Recobination}
\end{figure}

\begin{figure}[H]
    \centering
    \includegraphics[width=0.7\textwidth]{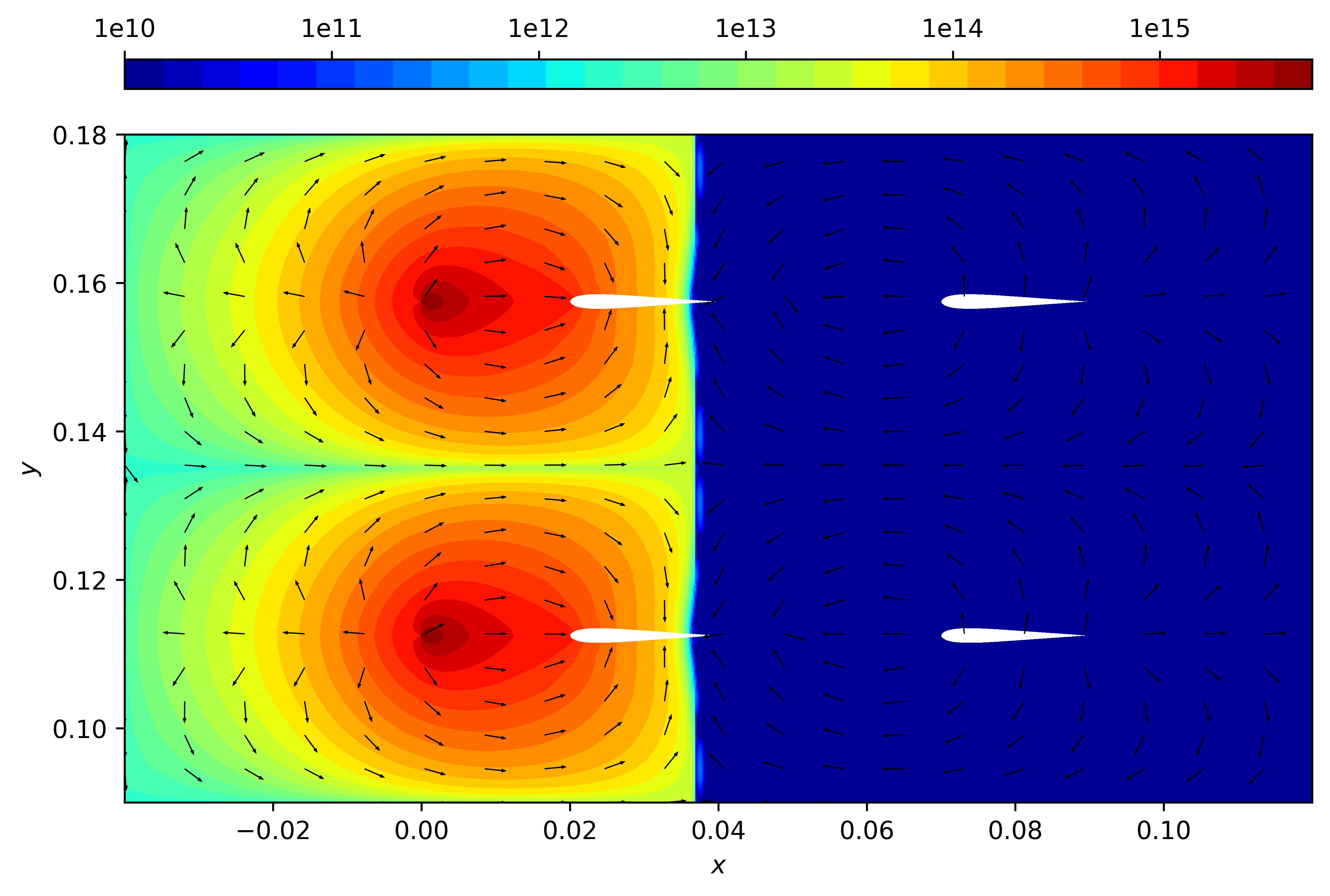}
    \caption[Positive Number Density]{Positive ions number density ($\text{m}^{-3}$) contours at $V_a = 14\,\text{kV}$. Only the two upper electrode pairs are shown. Arrows are tangent to the electric field.}
    \label{fig: PositiveDensity_14kV}
\end{figure}

\begin{figure}[H]
    \centering
    \includegraphics[width=0.7\textwidth]{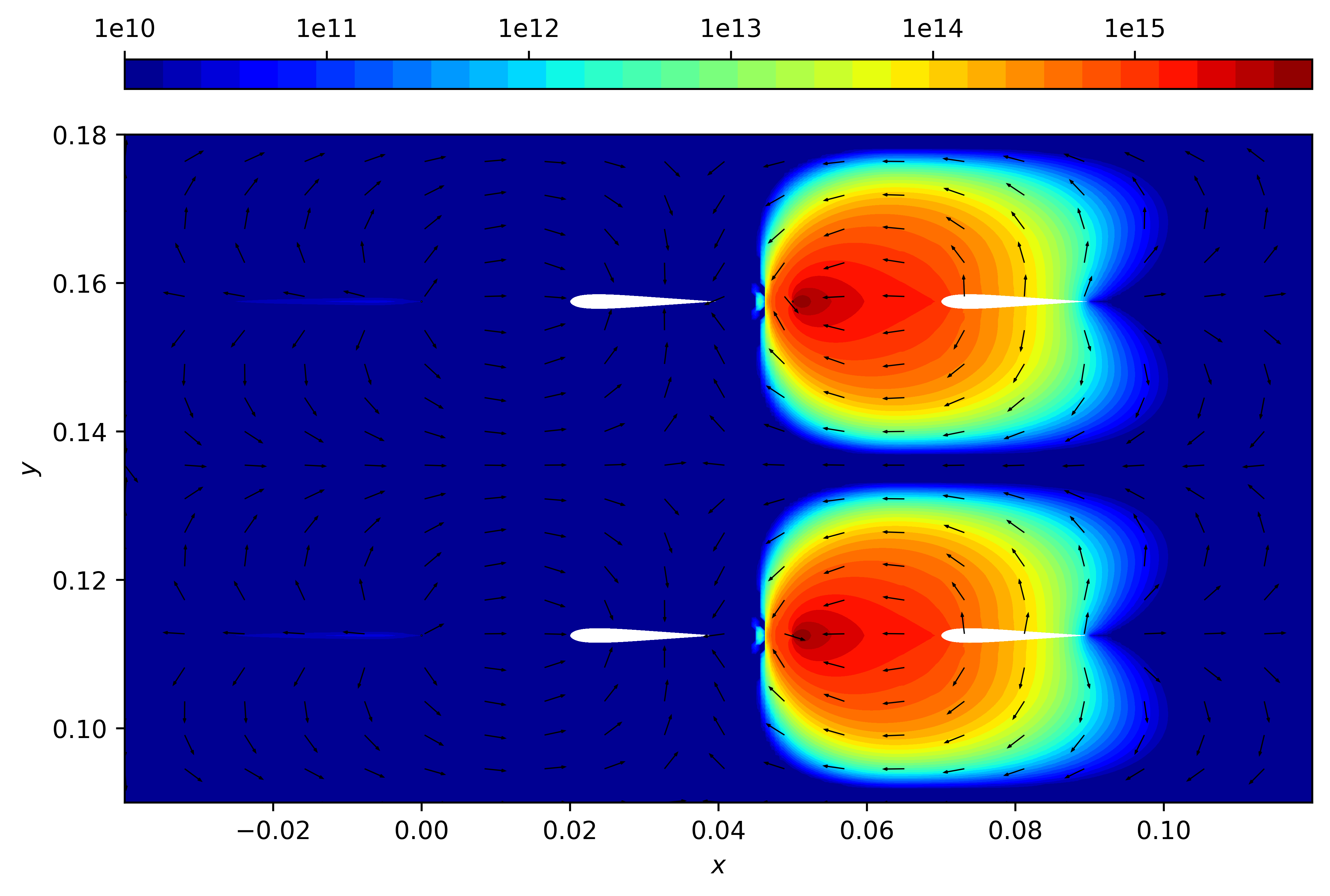}
    \caption[Positive Number Density]{Negative ions number density ($\text{m}^{-3}$) contours at $V_a = 14\,\text{kV}$. Only the two upper electrode pairs are shown. Arrows are tangent to the electric field.}
    \label{fig: NegativeDensity_14kV}
\end{figure}

The simulations of Secs.~\ref{subsec: single_stage_grid_convergence} and~\ref{sec:validation} have been carried out without recombination reactions on the r.h.s.\ of Eqs.~\ref{eq: DD_adim}. 
In the case of unipolar corona discharges, one of the two kinds of ions is confined close to the emitter electrode and therefore recombination reactions are neglected in the drift region~\cite{MonPlour,Calio}.
Recombination reactions could be relevant in a multistage configuration with a combination of positive and negative discharges, because a sufficiently high number density of both positive and negative ions could be present inside the drift region. 
In this Section we assess the effect of the recombination term on current and electrostatic thrust and show that it can be safely neglected for the configurations of interest in this work. We considered a thruster with two stages, each made of $4$ emitter-collector pairs, with $\Delta=45\,\text{mm}$ and gap $d=20\,\text{mm}$. The first stage operates as a positive corona, with the positive voltage applied to the emitters and the collectors grounded. The second stage operates as a negative corona, with the emitters grounded and the same positive voltage applied to the collectors. If the recombination term plays a role, it should be significant when the stages are close to each other and fades away when the distance, $\theta$, between stages increases. Indeed, in the limit $\theta\rightarrow \infty$, this would correspond to two isolated stages each operating as a unipolar discharge. Therefore we consider a pretty narrow distance $\theta$ equal to half of the gap $\theta=0.5\, d$. 
In Figure~\ref{fig: Recobination} current $I$ and electrostatic thrust density magnitude $T_E$ computed for different values of $R_i$ are shown. A reference value of the recombination coefficient at atmospheric pressure is~\cite{Bipolar} $R_i=1.8\times10^{-12}\,\text{m}^3\text{s}^{-1}$. Then we considered $R_i=0\,\text{m}^3\text{s}^{-1}$ and a value ten times greater than the reference one, $R_i=1.8\times10^{-11}\,\text{m}^3\text{s}^{-1}$. Differences in computed current and electrostatic thrust are negligible, hence allowing us to neglect the recombination term.\\
It is possible to understand why the effect of recombination is negligible by looking at the ions number density in Figs.~\ref{fig: PositiveDensity_14kV} and~\ref{fig: NegativeDensity_14kV}. In the former we show the distribution of positive ions, while in the latter the distribution of negative ions. The collectors of the first stage and the emitters of the second stage are grounded. For this reason, the electric field is almost null in the region between collectors of the first stage and emitters of the second. Therefore the positive ions basically stop near the trailing edge of the collectors of the first stage and the negative ions cannot move backward. As a consequence the positive and negative ions remain well confined in their respective stage. 
However, this conclusion might not be valid in case of strong adverse wind. In such a situation, the feedback effect of the fluid motion cannot be neglected anymore and ions can be transported by the fluid itself. The analysis should be carried out again in this scenario and will be the object of future works.

\subsection{Multi-stage Thruster with Wire Collectors}\label{sec: MultistagedGrid}

\begin{figure}[H]
    \centering
    \includegraphics[width=1\textwidth]{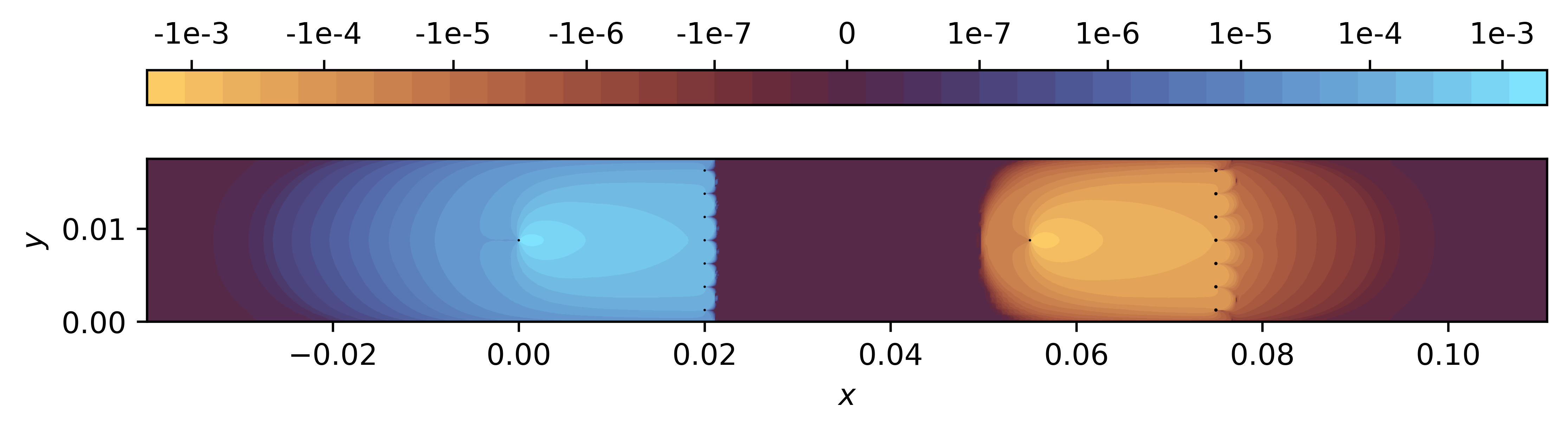}
    \caption[Space charge]{Computed charge density ($\text{C/m}^3$) at $V_a = 17\,\text{kV}$. }
    \label{fig: Bipolar_SpaceCharge}
\end{figure}

\begin{figure}[H]
    \centering
    \includegraphics[width=1\textwidth]{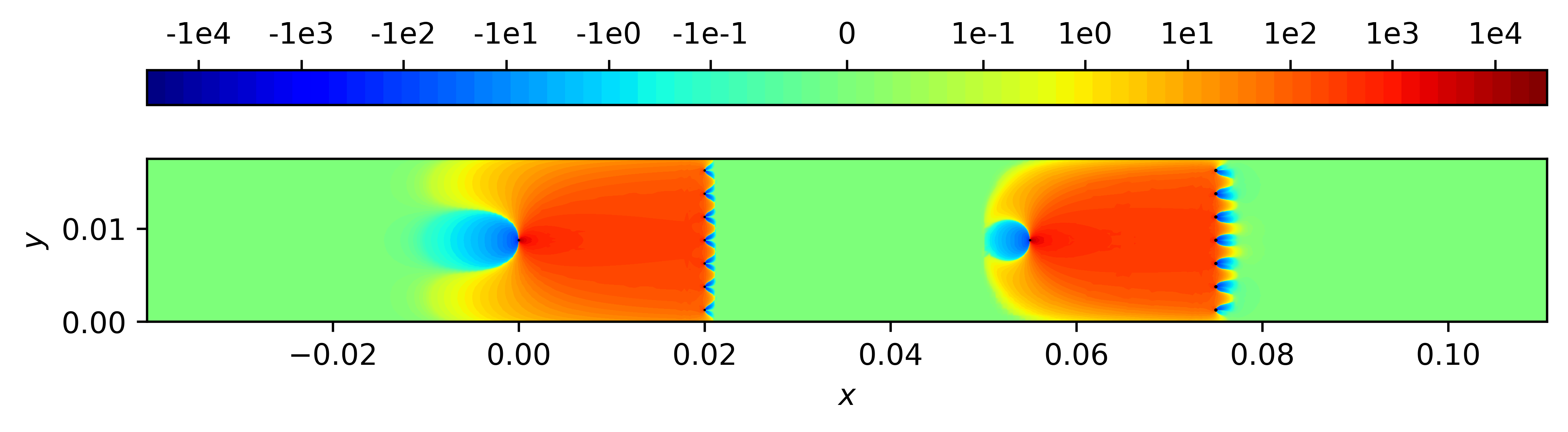}
    \caption[Space charge]{Computed electrostatic force density ($\mathrm{Nm^{-3}}$) at $V_a = 17\,\mathrm{kV}$. }
    \label{fig: Bipolar_Fx}
\end{figure}

\begin{figure}[H]
    \centering
    \includegraphics[width=0.53\textwidth]{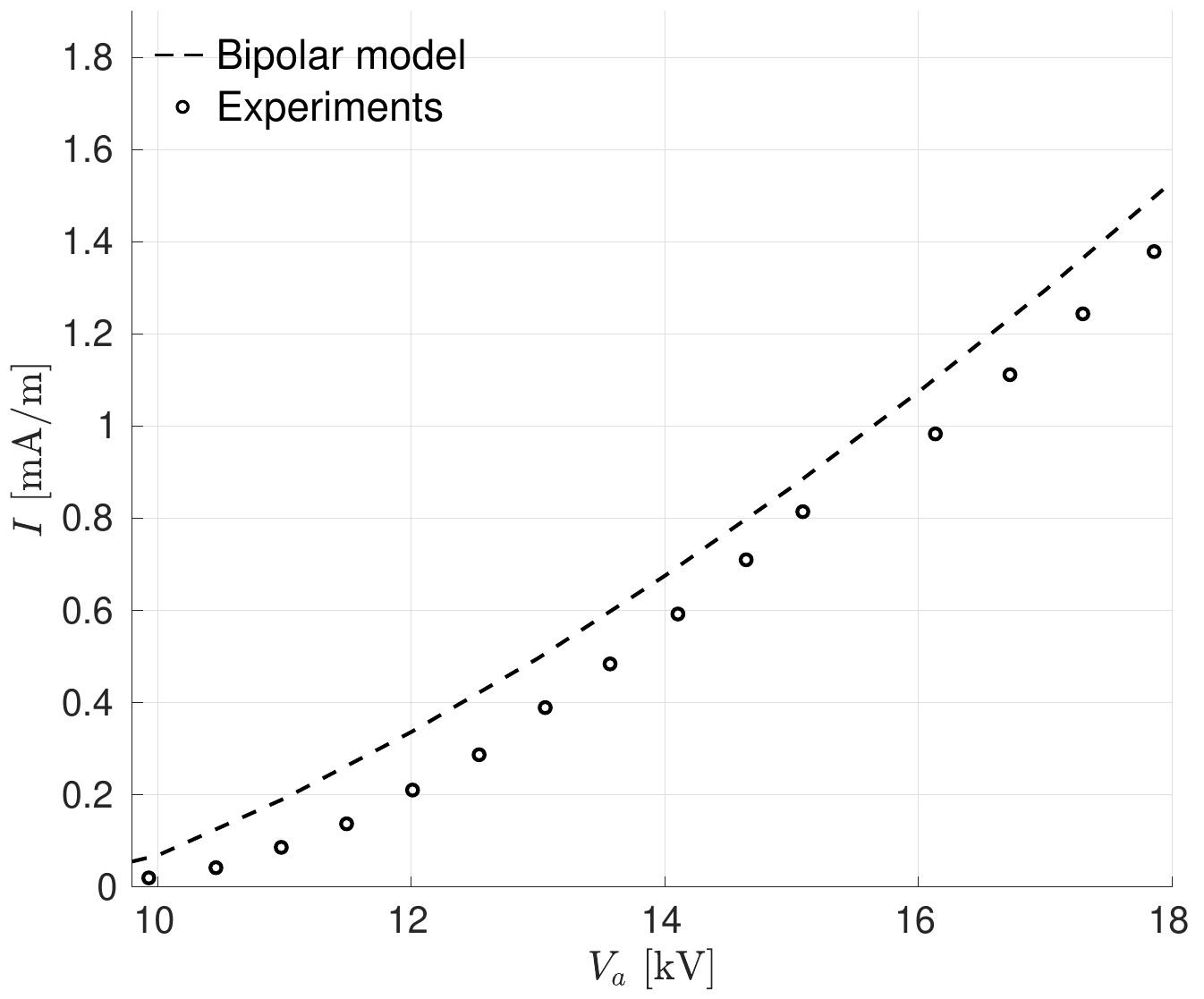}
    \caption[Space charge]{Computed current–voltage characteristics compared with experimental measurements for the two-stage thruster.}
    \label{fig: Current_Comparison}
\end{figure}

In this Section we study a two-stage configuration with wire collectors~\cite{Usuelli25,iprop_d1.3}.
The first stage of the thruster consists of a wire electrode with radius $r = 50\,\mu\text{m}$, biased at positive voltage $V_a$, separated by a $20\,\text{mm}$ air gap from a grounded grid composed of seven wires with radius $r = 50\,\mu\text{m}$. The second stage has opposite polarity configuration: a grounded wire electrode with radius $r = 50\,\mu\text{m}$ is separated by a $20\,\text{mm}$ air gap from a positively biased grid ($V_a$) composed of seven wires with radius $r = 100\,\mu\text{m}$. The stages are separated by a $35\,\text{mm}$ air gap. The injection law parameters for this configuration are given in Table~\ref{tab: gride_parameters}.
\begin{table}[H]
\centering
\begin{tabular}{l l l l l}
  \toprule
  \textbf{Variable} & \textbf{} & \textbf{Value} & &\textbf{Units}  \\
    \text{$n_{min}$} & & $10^{9}$                 & & $[\,\text{m}^{-3}\,]$    \\
    \text{$n_{ref}$} & & $10^{9}$                 & & $[\,\text{m}^{-3}\,]$    \\
    \text{$E_{ref}$} & & $9.8584\times 10^{4}$    & & $[\,\text{V}\,\text{m}^{-1}\,]$      \\
    \text{$\Bar{E}_{on}$} & & $1.478754\times10^{7}$     & & $[\,\text{V}\,\text{m}^{-1}\,]$   \\
\end{tabular}
\caption{Parameters used for two-stage configuration with wire collectors.}
\label{tab: gride_parameters}
\end{table}
Figure~\ref{fig: Bipolar_SpaceCharge} shows the charge density in the domain for $V_a=17\,\mathrm{kV}$. We notice the separation between positive and negative ions, with the region between them remaining nearly electrically neutral, because collectors of the first stage and emitter of the second stage are at the same voltage.
Figure~\ref{fig: Bipolar_Fx} shows the $x$-component of the electrostatic thrust density $\vec{f}_E$ for the same applied voltage. The results demonstrate that most of the electrostatic thrust is produced in the region between the emitter and the central collectors, whereas the peripheral collectors give only a minor contribution to the total thrust. 
The current-voltage curve for the thruster is shown in Fig.~\ref{fig: Current_Comparison}. A reasonable agreement with the experimental measurements is found; for applied voltages above $13\,\text{kV}$ the difference between our computations and experiments does not exceed $15\,\%$.


\section{Multi-stage Thruster: a Fluid Analysis}
\label{sec:perfeval}

\subsection{Experimental Validation of the Effective Thrust \texorpdfstring{$T_{EAD}$}{Tead}}

\begin{figure}[H]
    \centering
    \includegraphics[width=0.53\textwidth]{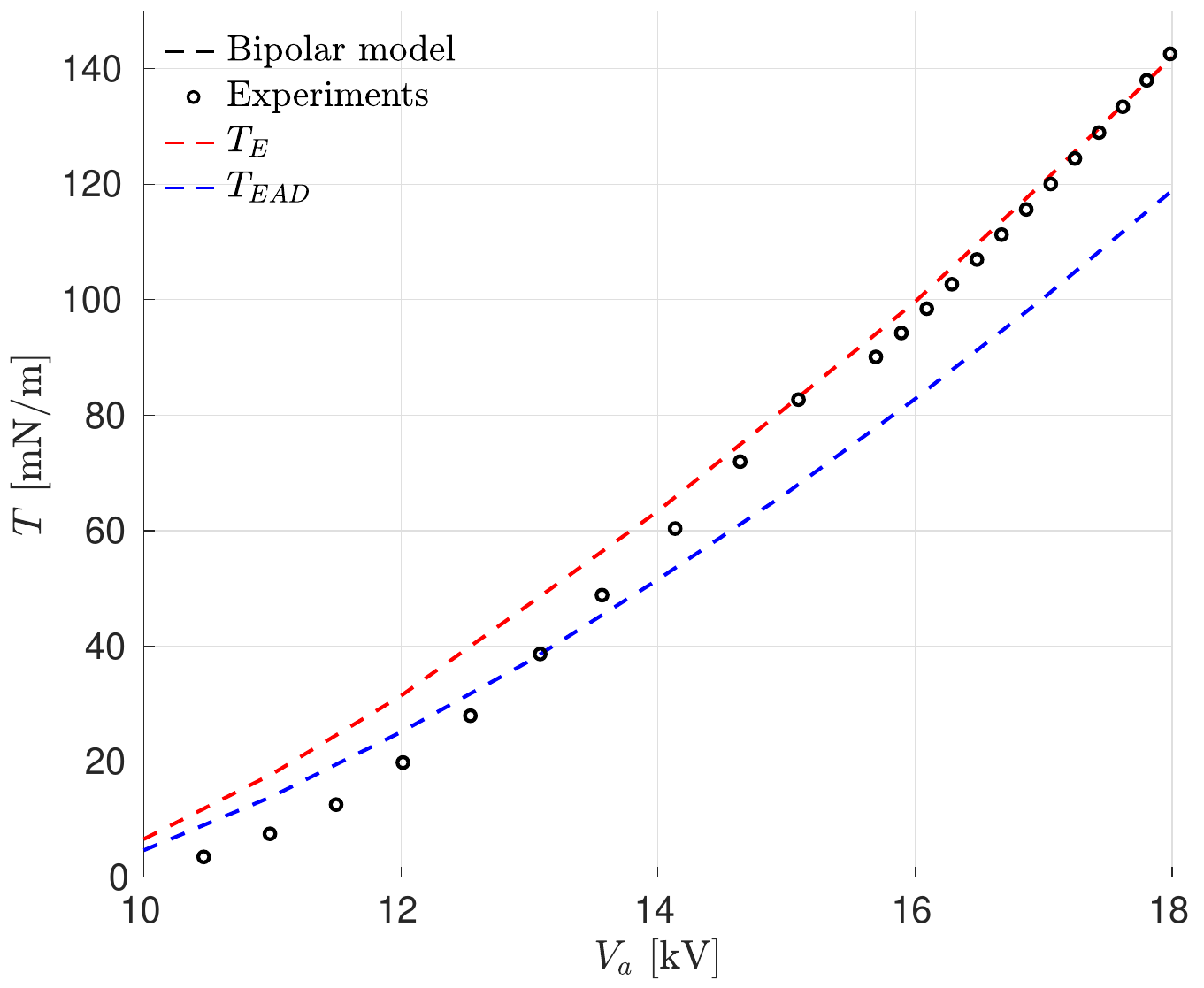}
    \caption[TV]{Simulated Thrust–voltage characteristics compared with experimental measurements for the double-stage thruster.}
    \label{fig: Thrust_Comparison}
\end{figure}

\begin{figure}[H]
    \centering
    \includegraphics[width=1\textwidth]{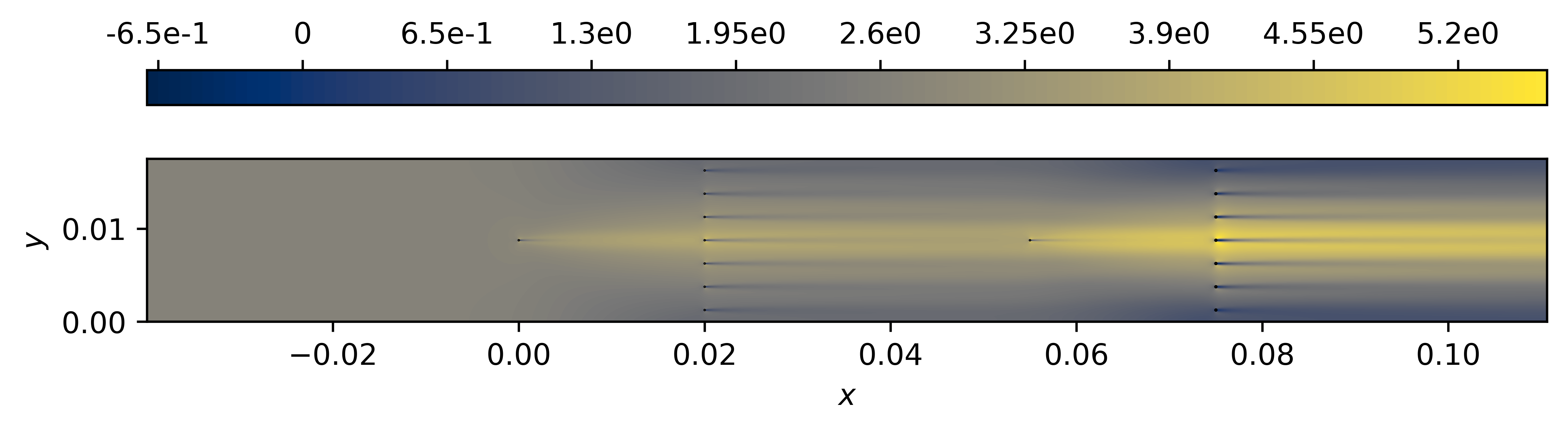}
    \caption[Vel]{Simulated velocity field at $V=17\, \text{kV}$ for the double-stage thruster.}
    \label{fig: VelocityXGrid}
\end{figure}

We consider the coupled Navier-Stokes plus corona simulation of the two-stage configuration with wire collectors analyzed in Section~\ref{sec: MultistagedGrid}. In Fig.~\ref{fig: Thrust_Comparison} we compare the x-components of electrostatic thrust $\vec{T}_E$ and effective thrust $\vec{T}_{EAD}$ with the experimental measurements. The computed effective thrust is always smaller than the electrostatic one, mainly because of the aerodynamic drag of the collectors. As should be expected, the difference between electrostatic and effective thrust increases with the voltage, because the higher the applied voltage, the higher the ionic wind velocity and therefore the drag. For an applied voltage higher than $12\,\mathrm{kV}$ the experimental results fall in between the numerically computed electrostatic thrust, $\vec{T}_E$ and effective thrust, $T_{EAD}$. The slope of the experimental thrust-voltage curve is higher for the experimental results than for the numerical ones. As a result, for lower applied voltage computed effective thrust and experimental thrust are in good agreement and for higher applied voltage effective thrust underestimates the experimental thrust, with the latter closer to computed electrostatic thrust.
The flow field along the horizontal direction is shown in Fig.~\ref{fig: VelocityXGrid}. We observe that the velocity is higher in the central region of the thruster. This is coherent with the distribution of the electrostatic thrust density $\vec{f}_E$ that peaks in the central region of the domain, as shown in Fig.~\ref{fig: Bipolar_Fx}.

\subsection{Effect of the Inter-stage Distance}

\begin{figure}[!ht]
    \centering
        \includegraphics[width=0.6\textwidth]{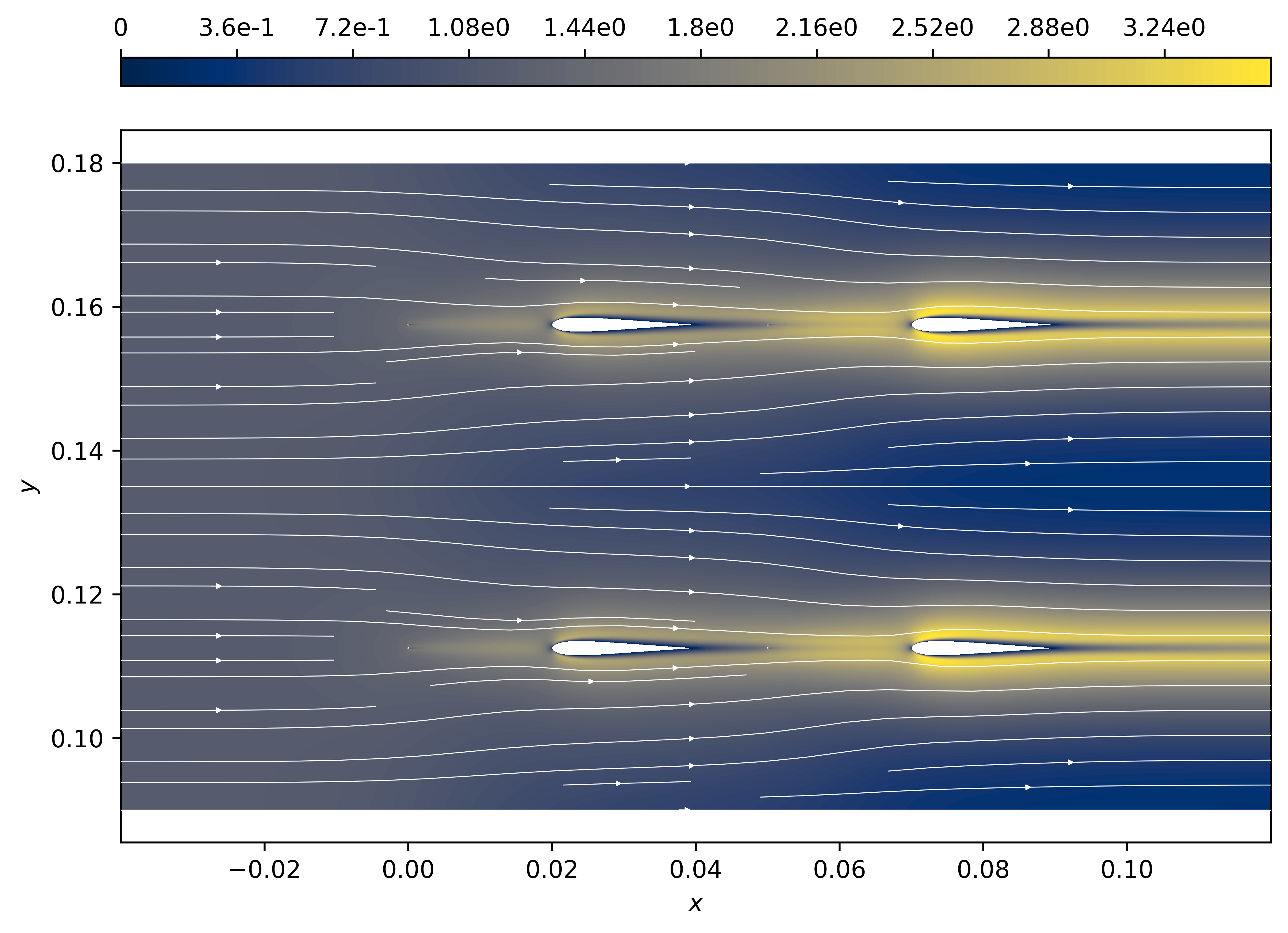}
    \caption[VelocityX]{Horizontal component of velocity with streamlines.}
    \label{fig: VelocityX}
\end{figure}

\begin{figure}[H]
    \centering
        \includegraphics[width=0.55\textwidth]{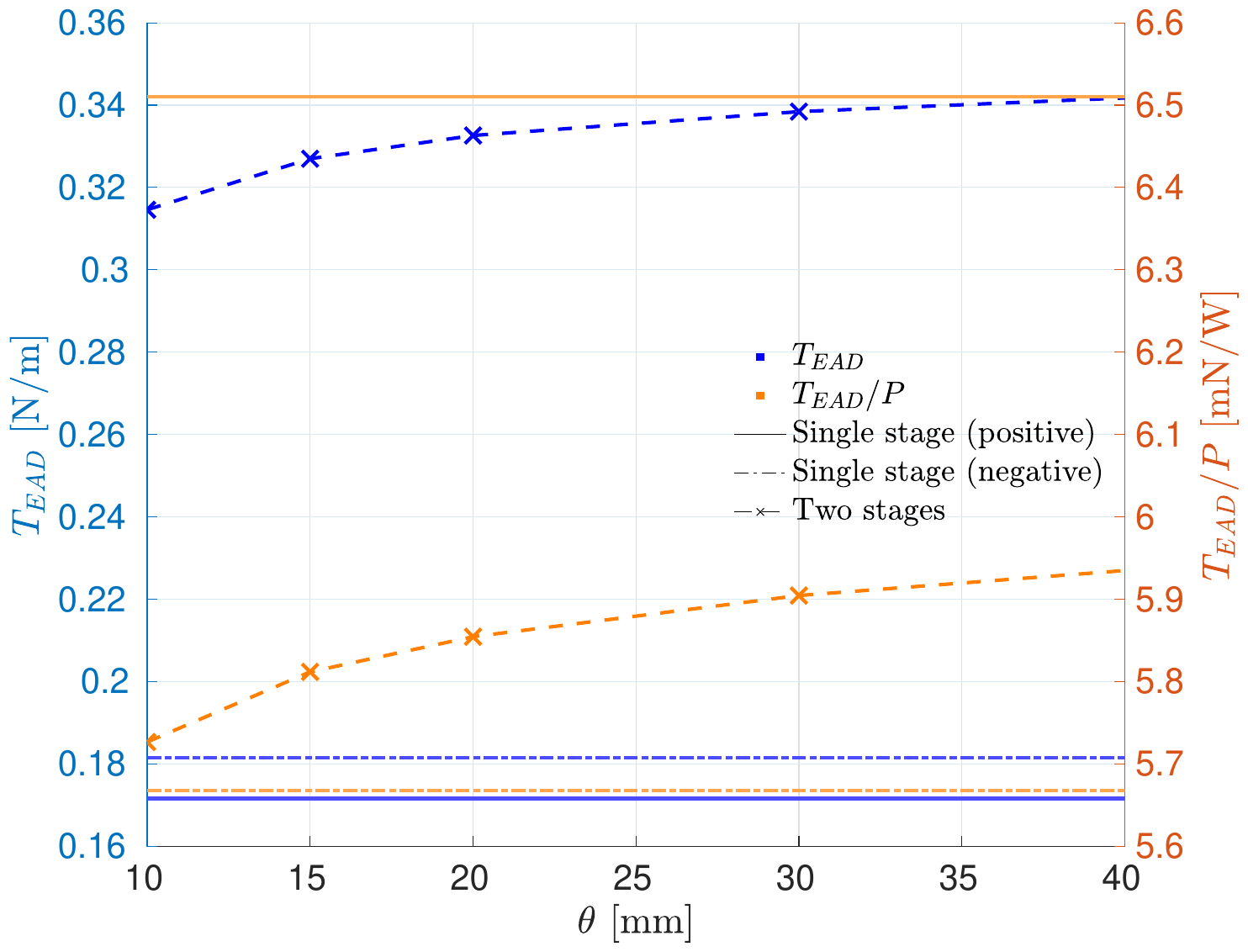}
    \caption{Two stage configuration. Thrust $T_{EAD}$ and thrust-to-power ratio as a function of the inter-stage distance $\theta$.}
    \label{fig:perfa}
\end{figure}

\begin{figure}[H]
    \centering
        \includegraphics[width=0.55\textwidth]{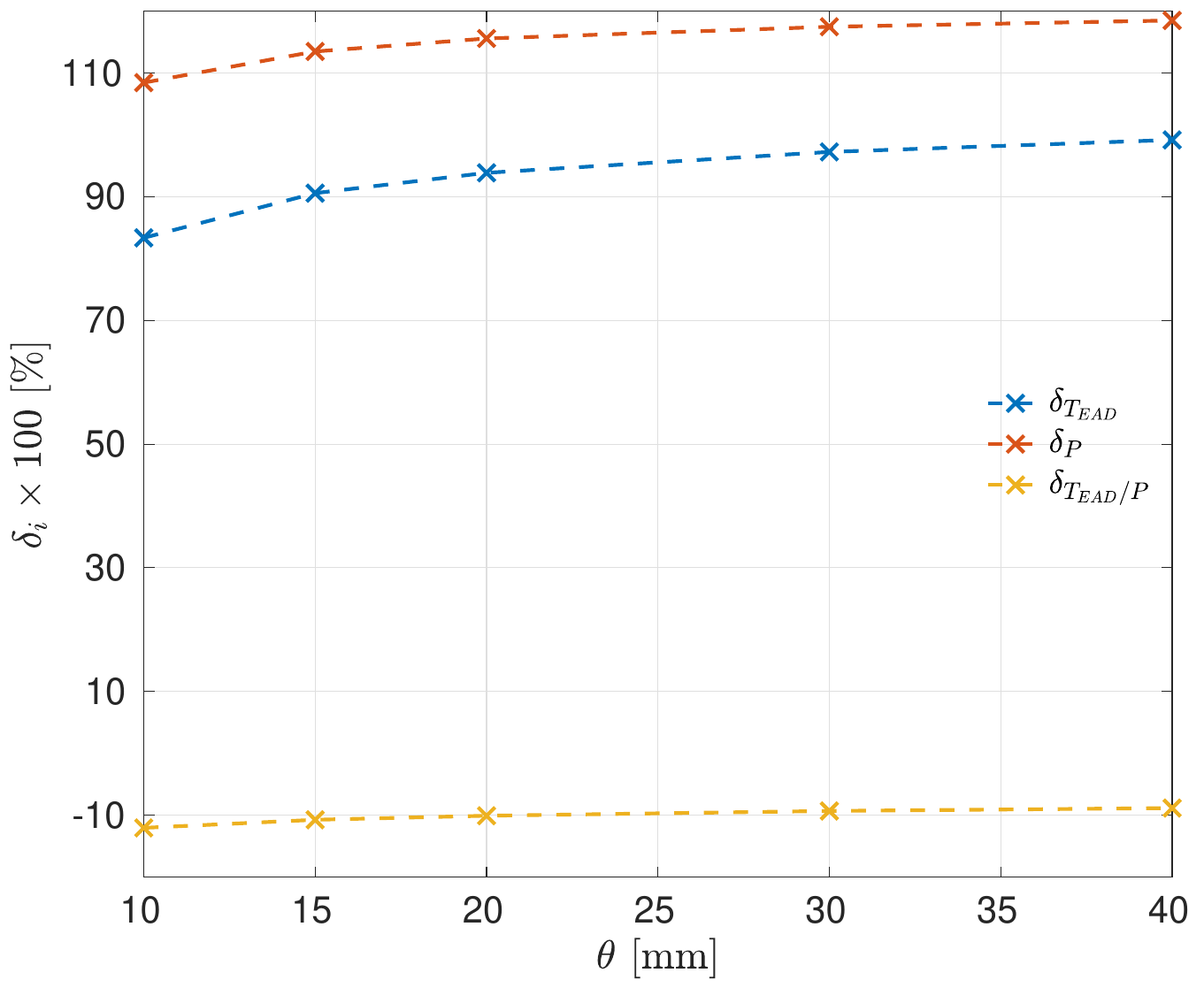}
        \caption{Relative variation of thrust, power and thrust-to-power ratio with respect to single stage positive corona.}
        \label{fig:perfb}    
\end{figure}

We study the effect of the inter-stage distance $\theta$ on the effective thrust $T_{EAD}$. We consider the configuration of Sec.~\ref{sec: Recombination}, with five values of the inter-stage distance $\theta= \left[10,\,15,\,20,\,30,\,40\right]\,\text{mm}$.\\
The horizontal component of the velocity field for the smallest inter-stage distance $\theta=10\,\text{mm}$ is shown in Fig.~\ref{fig: VelocityX}. One can notice that the velocity near the collectors of the second stage is higher than the velocity near the collectors of the first stage, as expected in a multi-stage thruster.
Figure~\ref{fig:perfa} shows the computed values of the effective thrust $T_{EAD}$ and of the thrust-to-power ratio as a function of the inter-stage distance. The thrust of the two stages configuration is always higher than the thrust of the single stage configuration and increases with the inter-stage distance $\theta$. Thrust growth is not unlimited however, but tends to an asymptotic limit. The latter condition corresponds to a configuration in which the electric field of one stage does not feel the influence of the electric field of the other stage. The increase in the generated thrust comes at the price of a higher power consumption; two-stage configurations produce a lower thrust-to-power ratio ($T_{EAD}/P$) with respect to the single stage configuration in positive corona. There are two reasons for this behavior: interference between the stages and the fact that negative corona thrusters have a lower efficiency as already pointed out in Sec.~\ref{sec:validation}.
The configuration with $\theta = 10\,\text{mm}$, for which there is the highest interference between the stages, is the least efficient.
Figure~\ref{fig:perfb} shows the relative variation $\delta_i$ of effective thrust, power and thrust-to-power ratio with respect to the single stage configuration in positive corona. The relative variation $\delta_i$ for each quantity of interest is defined as follows:
$$
\delta_i=\frac{i_{multi}-i_{single}}{i_{single}} \quad i=T_{EAD}, P, T_{EAD}/P
$$ 
It is interesting to note that, for all inter-stage distances, the power absorbed by the two stage thruster is always higher than the power absorbed by two separate stages, meaning that the interference between the electric fields is still felt at an inter-stage distance $\theta=40\,\mathrm{mm}$.
We observe that, in the two-stage configuration, the thrust-to-power ratio increases only slightly with the inter-stage distance and remains, on average, approximately $10\,\%$ lower than that of the corresponding single-stage configuration.

\section{Multiple Emitter Inception}\label{sec: MultipleEmitterInception}

In this Section we address the issue of the inception pattern when multiple emitters are considered. In Ref.~\cite{MultiInception}, the authors show how their model, based on the Kaptzov hypothesis, is able to capture the inception pattern of configurations with multiple emitters and one big wire as collector. Because of the way Kaptzov's assumption is enforced, the inception pattern cannot be automatically captured by the numerical solution and the authors are obliged to perform an iterative trial and error procedure~\cite{MultiInception}. Despite the correctness of the approach, it is computationally more efficient if the inception pattern of the emitters can be automatically captured by the numerical method. The injection law used in our model makes it possible, because the ion density at the emitters is imposed as a function of the local electric field that is self-consistently computed, while in Ref.~\cite{MultiInception} the electric field at the emitters is locked at the value dictated by Kaptzov's hypothesis and the local number density is computed as to satisfy the constraint on the electric field.
\begin{table*}[!ht]
\centering
\footnotesize

\begin{tabular}{|c|c|c|c|c|}
\hline
\text{Configuration} &
\text{$V_{E,exp}\,\mathrm{(kV)}$} &
\text{$V_{E,LMDF}\,\mathrm{(kV)}$} &
\text{$V_{E,Present}\,\mathrm{(kV)}$} &
\text{Inception state}\\
\hline

\multirow{2}{*}{\centering\arraybackslash
  \begin{tabular}{c}
    3 Emitters\\
    $d=45\,\mathrm{mm}$; $\Delta=60\,\mathrm{mm}$
  \end{tabular}
}
& $V^c_{E0}=10.5$ & $V^c_{E0}=11.5$ & $V^c_{E0}=11$ & \patIII{\emptyc}{\full}{\emptyc}\\ 
& $V^c_{E1}=12.5$ & $V^c_{E1}=17.50$ & $V^c_{E1}=12.25$ & \patIII{\full}{\full}{\full} \\

\hline

\multirow{2}{*}{\centering\arraybackslash
  \begin{tabular}{c}
    3 Emitters\\
    $d=100\,\mathrm{mm}$; $\Delta=60\,\mathrm{mm}$
  \end{tabular}
}
& $V^c_{E1}=13$ & $V^c_{E1}=14.75$ & $V^c_{E1}=13.25$ & \patIII{\full}{\emptyc}{\full}\\ 
& $V^c_{E0}=13.5$ & $V^c_{E0}=19.50$ & $V^c_{E0}=14$ & \patIII{\full}{\full}{\full} \\

\hline

\multirow{2}{*}{\centering\arraybackslash
  \begin{tabular}{c}
    3 Emitters\\
    $d=45\,\mathrm{mm}$; $\Delta=40\,\mathrm{mm}$
  \end{tabular}
}
& $V^c_{E0}=10.5$ & $V^c_{E0}=12.25$ & $V^c_{E0}=11$ & \patIII{\emptyc}{\full}{\emptyc}\\ 
& $V^c_{E1}=11$ & $V^c_{E1}=15.75$ & $V^c_{E1}=11.25$ & \patIII{\full}{\full}{\full} \\

\hline

\multirow{2}{*}{\centering\arraybackslash
  \begin{tabular}{c}
    3 Emitters\\
    $d=100\,\mathrm{mm}$; $\Delta=40\,\mathrm{mm}$
  \end{tabular}
}
& $V^c_{E1}=13$ & $V^c_{E1}=15$ & $V^c_{E1}=13.5$ & \patIII{\full}{\emptyc}{\full}\\ 
& $V^c_{E0}=23$ & $V^c_{E0}=29.75$ & $V^c_{E0}=15$ & \patIII{\full}{\full}{\full} \\

\hline

\multirow{3}{*}{\centering\arraybackslash
  \begin{tabular}{c}
    5 Emitters\\
    $d=45\,\mathrm{mm}$; $\Delta=20\,\mathrm{mm}$
  \end{tabular}
}
& $V^c_{E2}=13.5$ & $V^c_{E2}=17$ & $V^c_{E2}= 15.25$ & \patV{\full}{\emptyc}{\emptyc}{\emptyc}{\full}\\ 
& $V^c_{E0}=17.5$ & $V^c_{E0}=20.75$ & $V^c_{E0}=16.5$ & \patV{\full}{\emptyc}{\full}{\emptyc}{\full} \\
& $V^c_{E1}=21.5$ & $V^c_{E1}=25.75$ & $V^c_{E1}=16.5$ & \patV{\full}{\full}{\full}{\full}{\full} \\
\hline
\end{tabular}
\caption{Inception Pattern. Filled circles denote ignited electrodes. Blank ones non-ignited electrodes. $E0$ indicates the central emitter, while $Ei$ denotes the $i^{th}$ peripheral emitter pair, with the index increasing from the central emitter outward. $V^c_{Ei}$ is the applied voltage at which $i^{th}$ electrode pair is ignited.}
\label{tab: INCEPTION}
\end{table*}
To carry out the analysis we need to define an onset criterion. In analogy with Ref.~\cite{MultiInception} we considered an emitter ignited if the electric field on its surface is everywhere greater than the onset value $E_{on}$: the latter values being computed with Peek's formula~\cite{MultiInception}. Computations are carried out for two major configurations:
both have cylindrical emitters with $R_e=50\,\mathrm{\mu m}$ and cylindrical collector with $R_c=5\,\mathrm{mm}$. The first configuration has three emitters and one collector; the second one has five emitters and one collector. The first configuration in its turn has four sub-cases shown in Table~\ref{tab: INCEPTION} and the second one a unique sub-case, shown also in Table~\ref{tab: INCEPTION}. $\Delta$ is the distance between the centers of the emitters and $d$ is the face-to-face distance between emitters and collector, therefore the effective center-to-center distance between emitters and collector is equal to $R_e+R_c+d$.
The inception pattern is shown in Table~\ref{tab: INCEPTION}. 
Filled circles represent ignited electrodes and blank circles non-ignited electrodes. $V^c_{Ei}$ is the applied voltage at which the $i^{th}$ electrode pair is ignited. We compare our results with both experimental $V_{E,exp}$ and numerical $V^c_{E,LMDF}$ results of Ref.~\cite{MultiInception}. Results for the three emitters configuration are in good agreement with the experimental results. The ignition pattern is correctly captured and the computed ignition voltage is very close to the experimental value, except for the value of $V^c_{E0}$ in the fourth test-case. The agreement is less good for the five emitter configuration. The first ignition is correctly captured and the computed ignition voltage is $15.25\,\mathrm{kV}$ compared to an experimental value of $13.5\,\mathrm{kV}$, with a relative difference of $13\,\%$. However our computations predict that the other emitters ignite together at the same voltage of $16.5\,\mathrm{kV}$, differently from the experimental results and numerical results of Ref.~\cite{MultiInception}. 

\section{Conclusions}
The present work introduced a macroscopic bipolar model for corona discharges, coupled in a one-way fashion with the incompressible Navier-Stokes equations. One key feature of the proposed approach is to replace the complex ionization mechanism of corona discharges with an injection law that specifies, as a function of the local electric field, the amount of both positive and negative ions injected at the emitters. Due to its local nature, the model naturally allows for non uniform emission of ions around the emitter in non-symmetric configurations and correctly reproduces the inception pattern when multiple emitters are considered without a priori assumptions. At the same time, it retains the low computational cost typical of reduced approaches, while remaining significantly more efficient than full-plasma simulations. \\
The proposed model has been successfully applied to real EAD thrusters featuring both airfoil and wire collectors. Validation was carried out for single stage thrusters operating under both positive and negative polarities, as well as for a two-stage configuration with alternating polarity. The prediction of effective thrust was found to reproduce the experimental measurements, with an error of about $15\,\%$. \\
The CVFEM-SG method used to solve the corona discharge provides stable and accurate solutions, without requiring user-defined stabilization parameters. Furthermore, the C++ implementation based on the \textit{deal.II} library is computationally efficient, making the framework suitable for simulations even on standard laptops.

\section*{Acknowledgments}
This project has received funding from the European Union’s Horizon Europe Research and Innovation Programme under grant agreement No.\ 101098900. Views and opinions expressed are however those of the authors only and do not
necessarily reflect those of the European Union or European
Innovation Council and SMEs Executive Agency (EISMEA).
Neither the European Union nor the granting authority can be
held responsible for them.\\
Carlo de Falco acknowledges financial support from the project:
Prot.\ 2022TMW2PY - Transport phenomena in low dimensional structures: models, simulations and theoretical aspects. Funded by European Union, Next Generation EU, Missione 4-Componente 1-CUP D53D23005990006.
The simulations discussed in this work were, in part, performed on the HPC Cluster of the Department of Mathematics of Politecnico di Milano which
was funded by MUR grant Dipartimento di Eccellenza 2023-2027.

\section*{Data Availability Statement}
The data that support the findings of this study are available from the corresponding author upon reasonable request.

\appendix
\begin{appendices}
\numberwithin{equation}{section}
\section{Discretized Newton Method}\label{sec: discretized_newton}
In this section we want to give some implementation details on the global Newton method adopted for the Corona Problem.\\
Denoting with $z = (\varphi,n_p,n_n,E_{\Gamma_{e}})$, let us introduce the functional $\mathcal{F}(z)= \left\{
    \mathcal{R}_1(z),
    \mathcal{R}_2(z),
    \mathcal{R}_3(z),
    \mathcal{R}_4(z)\right\}$ in which $\mathcal{R}_i$ with $ i=1 ,2, 3$ is the residual of each component of the non linear system arising in the Corona problem \ref{eq: DD_adim} and $\mathcal{R}_4$ the residual for the equation of the electric field at the electrodes.

Recalling the notation introduced in section ~\ref{sec: NumericalMethods}, let define the following matrices:
\begin{align*}
    M_{ii} &= \vert C_i\vert  \\
    M^p_{ii}  &= \sum_{K_r \in \mathrm{K}(\mathbf{v}_i)} \langle n_{p} \rangle_{K_r} \vert C_i^r\vert  \\
    M^n_{ii}  &= \sum_{K_r \in \mathrm{K}(\mathbf{v}_i)} \langle n_{n} \rangle_{K_r} \vert C_i^r\vert  \\
    L_{ij} &= \sum_{K_r \in \mathrm{K}(\mathbf{v}_i)}\sum_{e_{jk} \in E(v_j)} \int_{\partial C_i} \vec{W}_{jk} \cdot {\hat{n}} \ dS \\
    L^{p}_{ij} &= \sum_{K_r \in \mathrm{K}(\mathbf{v}_i)} \sum_{e_{jk} \in E(v_j) \cap K_r} \langle n_{p} \rangle_{K_r} \int_{\partial C_i} \vec{W}_{jk} \cdot {\hat{n}} \ dS \\
    L^{n}_{ij} &= \sum_{K_r \in \mathrm{K}(\mathbf{v}_i)} \sum_{e_{jk} \in E(v_j) \cap K_r} \langle n_{n} \rangle_{K_r} \int_{\partial C_i} \vec{W}_{jk} \cdot {\hat{n}} \ dS \\
    K^{p}_{ij} &= \sum_{K_r \in \mathrm{K}(\mathbf{v}_i)}\sum_{e_{jk}\in E(v_j)} \frac{1}{Pe} B(2 a_{jk}) \int_{\partial C_i}\vec{W}_{jk}\cdot {\hat{n}} \ dS \\
    K^{n}_{ij}  &= K^{p}_{ij} (-\varphi_h)
\end{align*}
Where $\langle n_{p} \rangle_{K_r}$ and $\langle n_{n} \rangle_{K_r}$ are the average of $n_{p}$ and $n_{n}$ over the element $K_r$, respectively. The matrix $M_{ij}$ is the lumped mass matrix, $M^p_{ij}$ and $M^n_{ij} $ are the lumped mass matrices weighted by the average ions density. The matrix $L_{ij}$ is the CVFEM-SG Laplace matrix, $L^{p}_{ij} $ and $L^{n}_{ij}$ are the Laplace matrices weighted by the average ions. Finally $K^{p}_{ij} $ and $K^{n}_{ij}$ are the CVFEM-SG Drift-Diffusion matrices. We also define $M^{\Gamma_{e}}_{ii}$ as the lumped boundary mass matrix at the emitter boundary $\Gamma_e $. For a boundary vertex $\mathbf{v}_i$ whose neighboring boundary vertices are $\mathbf{v}_l$ and $\mathbf{v}_k$, the corresponding entry is given by:
$$
M^{\Gamma_{e}}_{ii}= \frac{h_{il}+h_{ik}}{2}
$$

After discretizing the governing equations with the CVFEM-SG method, we obtain, at each Newton iteration $k$, a set of residual vectors whose $i$-th components correspond to the residuals associated with the mesh vertex $\mathbf{v}_i$:
\begin{align*}
    \mathcal{R}_1(z^{(k)} )_i &=  L_{ij} \, \varphi_j^{(k)} -  M_{ij} \, n_{p,\,j}^{(k)} +  M_{ij} \, n_{n,\,j}^{(k)} \quad \forall \mathbf{v}_i \in V(\Omega) \\[8pt]
    \mathcal{R}_2(z^{(k)} )_i &= 
    \begin{cases}
        K^{p}_{ij} \, n_{p,\,j}^{(k)} + \frac{R^p}{Pe}  M_{ij} \left( n_{p,\,j}^{(k)} \, n_{n,j}^{(k)} - n_{min}^2\right)\quad \forall \mathbf{v}_i \in V(\Omega)  \\[8pt]
        n_{p,\,i}^{(k)} - f_p(E^{(k)}_{\Gamma_{e},\,i}) \quad \forall i \in V(\Gamma_{e}) \\[8pt]
    \end{cases} 
    \\[8pt]
    \mathcal{R}_3(z^{(k)} )_i &= 
    \begin{cases}
        \vspace{0.2cm}
        K^{n}_{ij} \, n_{n,\,j}^{(k)} + \frac{R^n}{Pe}  M_{ij} \left( n_{p,\,j}^{(k)} \, n_{n,\,j}^{(k)} - n_{min}^2\right) \quad \forall \mathbf{v}_i \in V(\Omega)   \\[8pt]
        n_{n,\,i}^{(k)} - f_n(E^{(k)}_{\Gamma_{e},\,i}) \quad \forall \mathbf{v}_i \in V(\Gamma_{e})  \\[8pt]
    \end{cases} 
    \\[8pt]
    \mathcal{R}_4(z^{(k)} )_i &= 
        M^{\Gamma_{e}}_{ii} \, E_{\Gamma_{e}, \,i}^{(k)} - \mathcal{R}_1(z^{(k)} )_i \quad \forall \mathbf{v}_i \in V(\Gamma_{e})
\end{align*}
where $f_p(E_{\Gamma_{e}})$ and $f_n(E_{\Gamma_{e}})$ are the functions defined in eq. \ref{eq: BC_ion} expressing the ions injection.
The applied voltage on the electrodes is set at the beginning of the Newton iteration and the increment of corresponding degree of freedom is enforced to be zero.\\
Thus the linear system to be solved at each inner iteration $k$ is:
 
\begin{equation*}
\!\!\!\!\!\!
\left(
\begin{array}{ c |c c | c  c |c }
L   & -M & -M & M  & M & 0 \\[6pt] \hline
L^p& K^p + \frac{R^p}{Pe} M^n & 0 & \frac{R^p}{Pe} M^p & 0 & 0 \\[6pt]
0 & 0 & I & 0 & 0 & -D_{iag}(\frac{\partial f_p(E^{(k)}_{\Gamma_{e}})}{\partial E_{\Gamma_e}})\! \\[6pt] \hline
L^n & \frac{R^n}{Pe} M^n &  0& K^n + \frac{R^n}{Pe} M^p & 0&0 \\[6pt]
0 & 0 & 0 & 0 & I & -D_{iag}(\frac{\partial f_n(E^{(k)}_{\Gamma_{e}})}{\partial E_{\Gamma_e}})\! \\[6pt] \hline
-L & 0 & M & 0 & -M & M^{\Gamma_{e}} \\[6pt]
\end{array}
\right) \!
\begin{pmatrix}
\delta \varphi^{(k)}_{\Omega} \\[6pt] \hline 
\delta {n}^{(k)}_{p,\Omega} \\[6pt]
\delta n^{(k)}_{p,\Gamma_{e}} \\[6pt] \hline
\delta n^{(k)}_{n,\Omega} \\[6pt]
\delta n^{(k)}_{n,\Gamma_{e}} \\[6pt] \hline
\delta E^{(k)}_{\Gamma_{e}} \\[6pt]
\end{pmatrix}
\!= \! - \!
\begin{pmatrix}
\mathcal{R}_1(z^{(k)} )_{\Omega} \\[6pt] \hline 
\mathcal{R}_2(z^{(k)} )_{\Omega} \\[6pt]
\mathcal{R}_2(z^{(k)} )_{\Gamma_{e}} \\[6pt] \hline
\mathcal{R}_3(z^{(k)} )_{\Omega} \\[6pt]
\mathcal{R}_3(z^{(k)} )_{\Gamma_{e}} \\[6pt] \hline
\mathcal{R}_4(z^{(k)} )_{\Gamma_{e}} \\[6pt]
\end{pmatrix}
\end{equation*}

Where $I$ is the identity matrix of appropriate size, $D_{iag}(\cdot)$ indicates a diagonal matrix with the given argument on the diagonal element, and
$$\frac{\partial f_p(E^{(k)}_{\Gamma_{e},\,i})}{\partial E_{\Gamma_e}} =  \frac{n_{ref}}{E_{ref}}\ \exp \left({\dfrac{E_{\Gamma_{e}, i}^{(k)} - E_{on}}{E_{ref}}}\right), \qquad \frac{\partial f_n(E^{(k)}_{\Gamma_{e},\,i})}{\partial E_{\Gamma_e}} =- \frac{n_{ref}}{E_{ref}}\ \exp \left({\dfrac{-E_{\Gamma_{e}, i}^{(k)} - E_{on}}{E_{ref}}}\right).$$
are the derivative with respect to the electric field of the injection law \ref{eq: BC_ion}.
\end{appendices}
\bibliographystyle{elsarticle-num}
\bibliography{bibliography} 
\end{document}